\documentclass[a4paper,fleqn,usenatbib]{mnras}

\usepackage{newtxtext,newtxmath}
\usepackage[T1]{fontenc}
\usepackage{ae,aecompl}

\usepackage{graphicx}	% Including figure files
\usepackage{amsmath}	% Advanced maths commands
\title[Semi-Transparent Shear Turbulence ]{Semi-Transparent Shear Turbulence in Hot Jupiter Atmospheres}

\author[K. Menou]{
Kristen Menou$^{1, 2, 3}$
\\
$^{1}$  Physics \& Astrophysics Group, Dept.   of  Physical  \&  Environmental  Sciences,  University  of  Toronto  Scarborough,\\   1265 Military Trail, Toronto, Ontario, M1C 1A4, Canada \\
$^{2}$ Dept.  of Astronomy \& Astrophysics, University of Toronto.
50 St.  George Street, Toronto, Ontario, M5S 3H4, Canada 
\\
$^{3}$ Dept.  of Physics, University of Toronto.
60 St George Street, Toronto, Ontario, M5S 1A7, Canada \\
}

\date{Accepted XXX. Received YYY; in original form ZZZ}

\pubyear{2015}

\begin{document}
\label{firstpage}
\pagerange{\pageref{firstpage}--\pageref{lastpage}}
\maketitle

% Abstract of the papers
\begin{abstract}
Turbulent transport driven by secular shear instabilities can lead to enhanced vertical mixing in hot Jupiter atmospheres, impacting their cloudiness, chemistry and overall vertical structure. We discuss the turbulent regime expected and evaluate theoretical uncertainties on the strength of the vertical mixing (i.e, $K_{\rm zz}$ values). We focus our work on three well-studied hot Jupiters with a hierarchy of atmospheric temperatures: HD189733b ($T_{\rm eq} \simeq 1200$K), HD209458b ($T_{\rm eq} \simeq 1450$K) and Kepler7b ($T_{\rm eq} \simeq 1630$K). $K_{\rm zz}$  uncertainties are large. They are dominated by i) the poorly understood magnitude of turbulent transport and ii) the semi-transparent nature of shear turbulence near the planetary photosphere.  Using a specific Moore-Spiegel instability threshold, we infer that the cooler HD189733b is not subject to enhanced mixing from semi-transparent shear turbulence while the daysides of the hotter Kepler7b and (marginally so) HD209458b are.  Enhanced vertical mixing is generally expected to manifest on hot enough exoplanets, with $T_{\rm eq} > 1500-1600$K.  On a given planet, day and night $K_{\rm zz}$ profiles can differ by an order of magnitude or more. Vertical mixing is slightly favoured in equatorial regions, where the atmospheric zonal shear is strongest. In all three planetary cases studied, momentum feedback on the atmospheric mean flow is minor to negligible. 
\end{abstract}


%%%%%%%%%%%%%%%%%%%%%%%%%%%%%%%%%%%%%%%%%%%%%%%%%%

%%%%%%%%%%%%%%%%% BODY OF PAPER %%%%%%%%%%%%%%%%%%

\section{Introduction}

Transiting exoplanet science is expected to enter an era of precision spectroscopy with the advent of JWST: high quality transmission and emission spectra of warm-to-hot planets ($T_{\rm eq} \geq 600$), from super-Jupiters down to the Neptune-mass regime, is on the horizon \citep{2016ApJ...817...17G,2016ApJ...833..120R,2017A&A...600A..10M,2018haex.bookE..85B, 2018haex.bookE.116P}. As this step forward in our ability to observationally characterize hot exoplanet atmospheres approaches,  it has also become clear that our ability to interpret this rich dataset is currently limited by the quality of our atmospheric modeling tools  \citep[e.g.][]{2019A&A...623A.161C, 2019ARA&A..57..617M,2020A&A...636A..66P, 2020ApJ...893L..43M}.

One particularly challenging aspect of atmospheric dynamics concerns the magnitude of vertical mixing and its role in preventing the otherwise inevitable gravitational settling of condensates \citep{2015AREPS..43..509H, 2018ApJ...860...18P, 2019AREPS..47..583H}. Indeed, vertical mixing can strongly impact the chemistry, cloud properties and even the entire atmospheric structure via its coupling to cloud radiative feedbacks. Recent illustrative examples of the crucial role that can be played by cloud radiative feedbacks, and the underlying dependence on the magnitude of vertical mixing, can be found in \cite{2019MNRAS.488.1332L,2019ApJ...872....1R,2021ApJ...908..101R}. Current observations already provide ample evidence for the role of clouds in shaping hot Jupiter observables \citep{2016Natur.529...59S, 2016ApJ...828...22P, 2019ApJ...881..152K}.

In this context, various efforts to better characterize the nature and magnitude of vertical transport in hot Jupiter atmospheres have been pursued. The vertical transport is often modeled as a diffusive process, with a magnitude quantified by an eddy viscosity ($K_{\rm zz}$) whose value varies throughout the atmosphere, in particular in the vertical. \cite{2013A&A...558A..91P} have inferred a  $K_{\rm zz}$ vertical profile for the hot Jupiter HD209458b based on the mixing pattern found in a 3D global circulation model. \cite{2018ApJ...866....1Z,2018ApJ...866....2Z} have developed a general theoretical framework to understand and describe the global vertical tracer mixing in planetary atmospheres. \cite{2019ApJ...881..152K}  built on this work to provide estimates of $K_{\rm zz}$ as a function of planetary effective temperature, chemical timescales and other relevant planetary parameters.    

One specific mechanism that has been proposed to enhance, and possibly dominate, vertical transport in Hot Jupiter atmospheres is the turbulence that results from double-diffusive (secular) shear instabilities \citep{2019MNRAS.485L..98M}.  Double-diffusive shear instabilities rely on efficient heat transport to neutralize the stabilizing role of thermal stratification, allowing some unstable shear modes to become unstable despite the strong background stratification. The analysis in \cite{2019MNRAS.485L..98M} assumed an optically-thick regime,  however.  In the present work, we improve on this original proposal by clarifying the semi-transparent nature of hot Jupiter atmospheres (with fluid perturbations that are optically-thin on small enough scales) and by better quantifying the expected magnitude of vertical transport, accounting for dominant sources of uncertainties.  Our work can be viewed as complementary to that of \cite{2019ApJ...881..152K}, which provides estimates of the $K_{\rm zz}$ values expected in the absence of turbulent transport from secular shear instabilities.

The plan of this article is as follows. In \S2, we present the global circulation models for three Hot Jupiters that are used in our analysis. In \S3, we discuss what we consider to be the three dominant sources of uncertainties affecting our estimates of $K_{\rm zz}$ values for the secular shear mechanism. We conclude in \S4.

\section{Hot Jupiter Global Circulation Models} 

\subsection{Three Hot Jupiter Models}

\begin{table*}
	\centering
	\caption{Planet-Specific Model Parameters}
	\begin{tabular}{ccccc} % four columns, alignment for each
		\hline
		Parameter & HD209458b & HD189733b & Kepler7b & Brief Description\\
		\hline
		GA & 8.0 & 22.0 & 4.14 & Gravitational Acceleration (m/s$^2$)\\ 
		SIDEREAL\_DAY & $285120$ &  $190080$ & $421690$ & Spin rotation period (s) \\ 
		RADIUS & $10^8$ & $8 \times 10^7$  & $1.12 \times 10^8 $ & Planet Radius (m) \\ 
		PSURF & $10^7$ &  $10^7$& $10^7$& Surface pressure (Pa)\\ 
		OOM & 5 &  5 & 5 & Pressure dynamic range: log10(PSURF/P$_{\rm top}$) \\ 
		AKAP & $0.286$ & $0.286$  & $0.286$ & Ratio of gass constant to specific heat\\ 
		GASCON & $3523$ &  $3523$ &  $3523$& Atmospheric gas constant\\ 
		TGR & $2700$ & $2700$  &  $2700$& Initial temperature (K)\\
		$T_{\rm eq}$ & 1450 & 1200 &  1630 & Equilibrium temperature (K)\\
		\hline
	\end{tabular}
    \label{table:one}
\end{table*}

\begin{figure}
	%\includegraphics[width=\columnwidth]{UZ_profile_noVD.pdf}
	\includegraphics[width=\columnwidth]{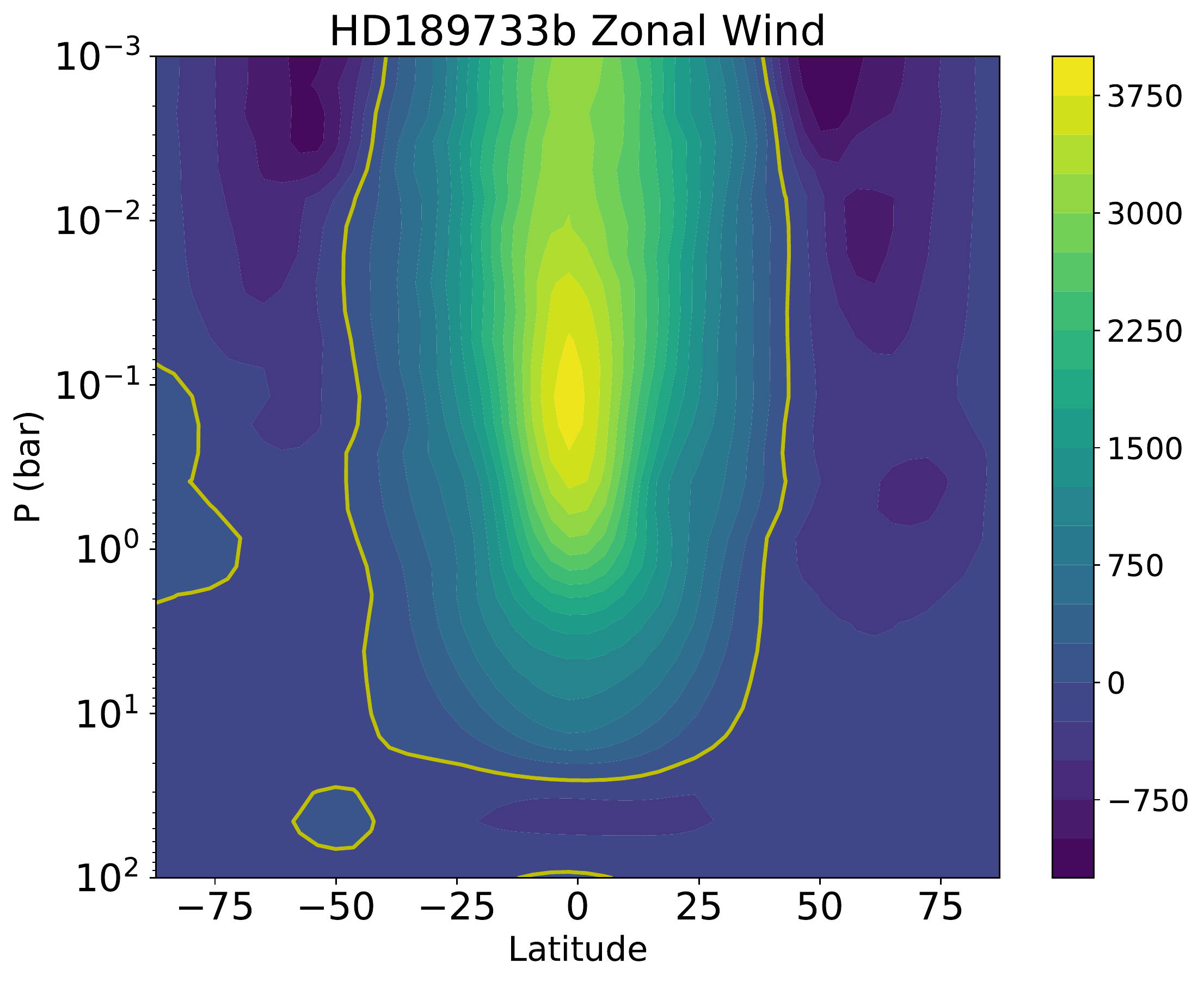}
	\includegraphics[width=\columnwidth]{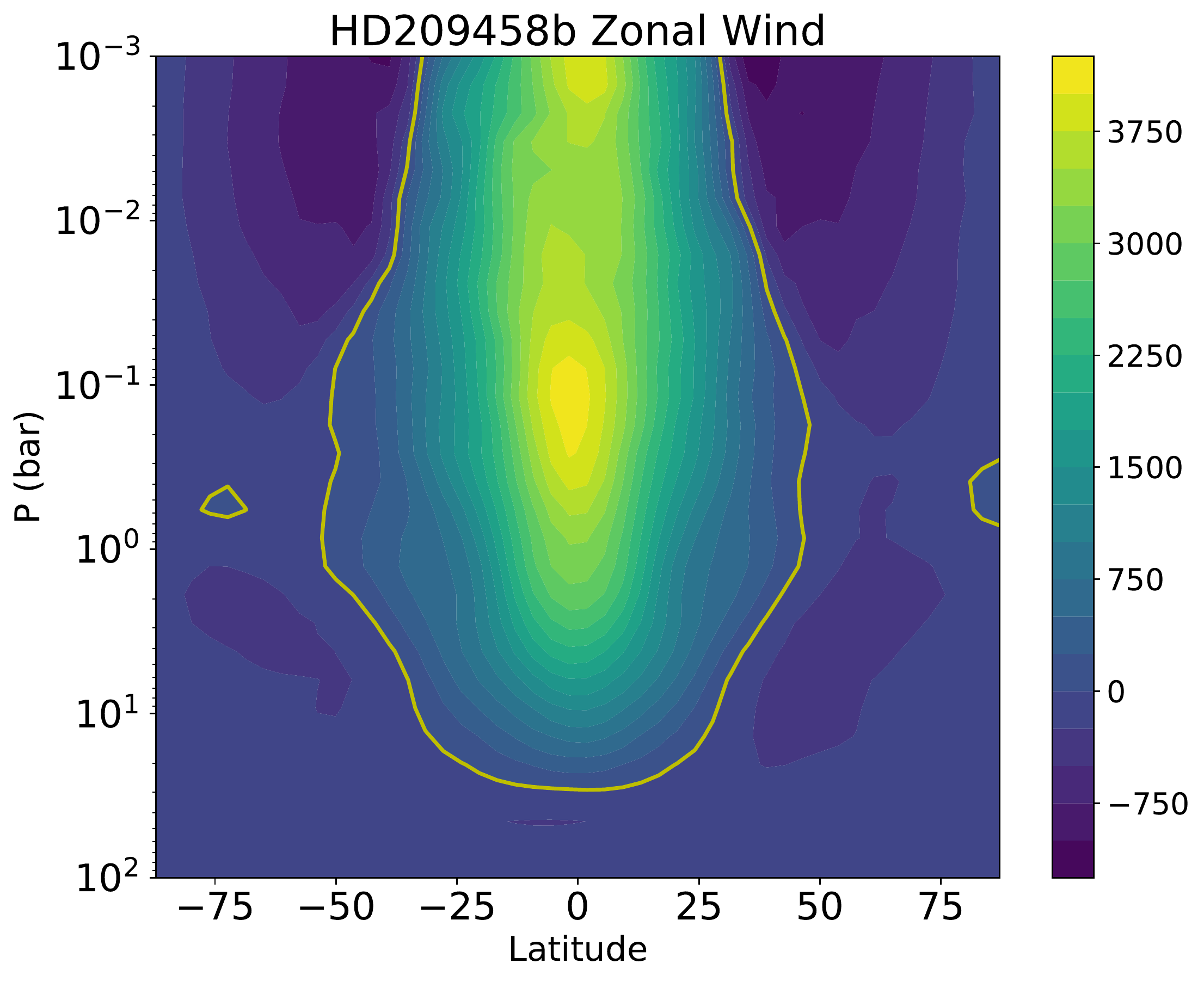}
	\includegraphics[width=\columnwidth]{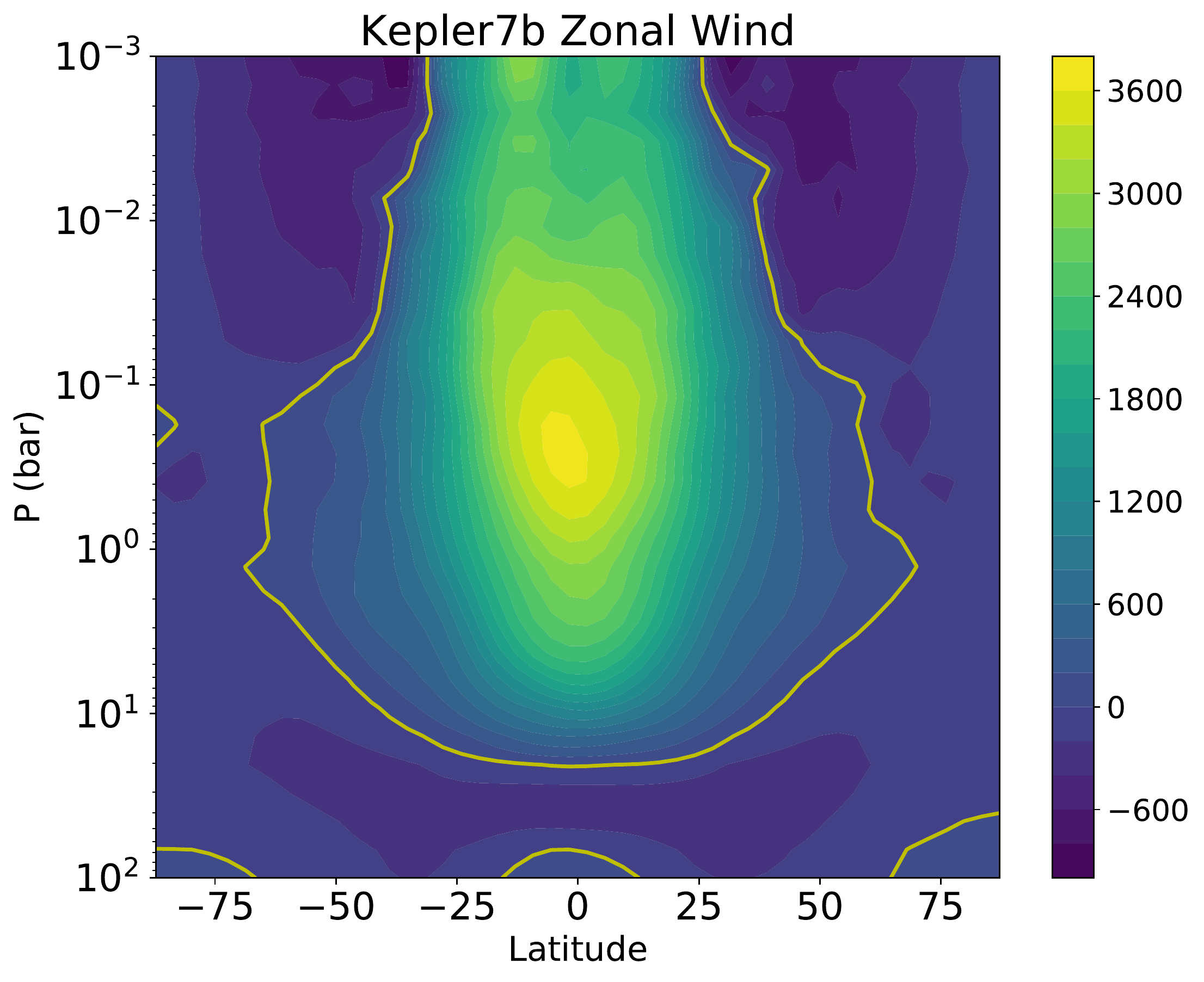}
    \caption{Zonally-averaged zonal wind profile (in m/s) at planet day 600 in our T31L30 models of HD189733b (top panel), HD209458b (middle) and Kepler7b (bottom). }
    \label{fig:zero1}
\end{figure}

\begin{figure}
	%\includegraphics[width=\columnwidth]{UZ_profile_noVD.pdf}
	\includegraphics[width=\columnwidth]{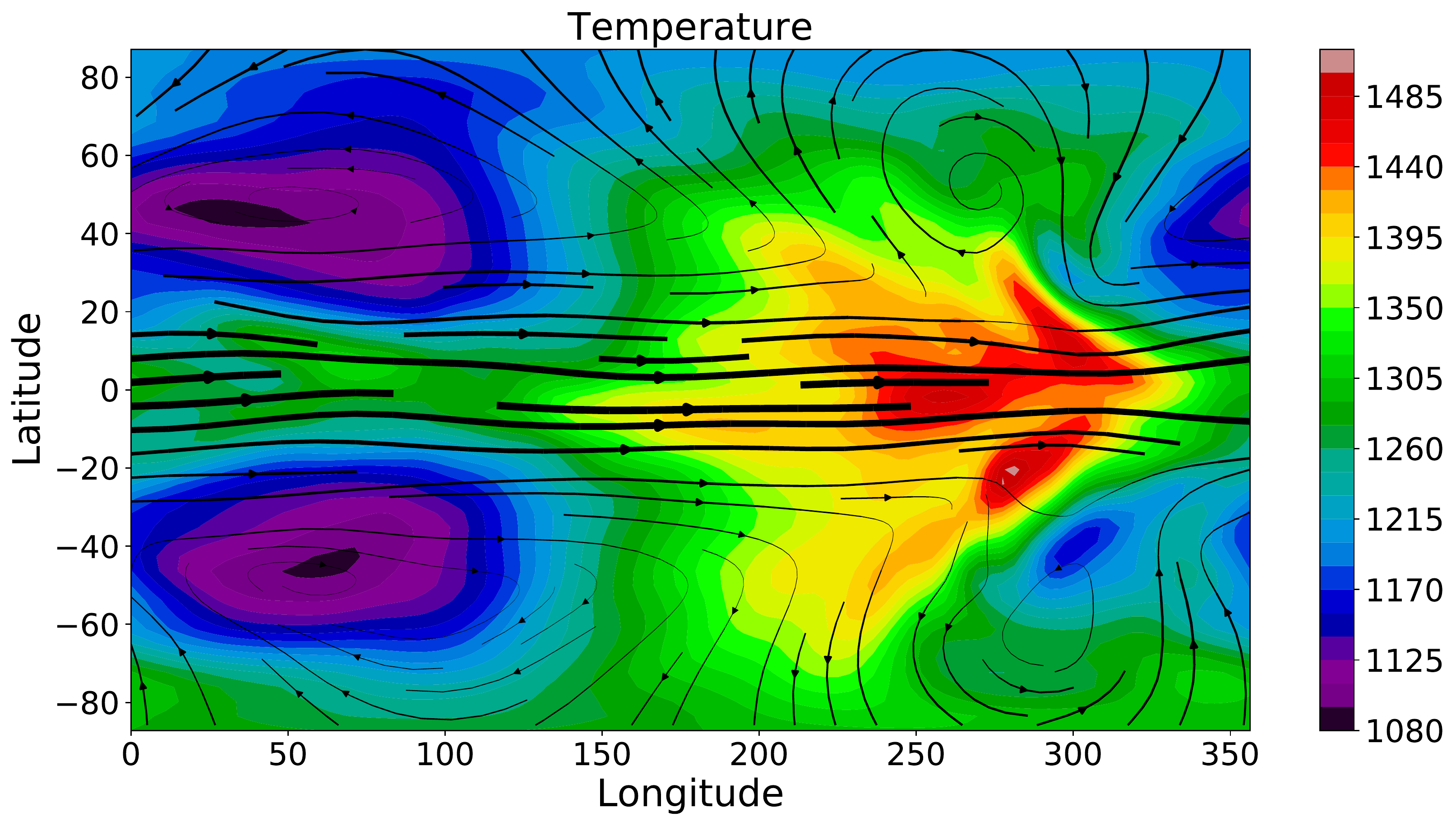}
	\includegraphics[width=\columnwidth]{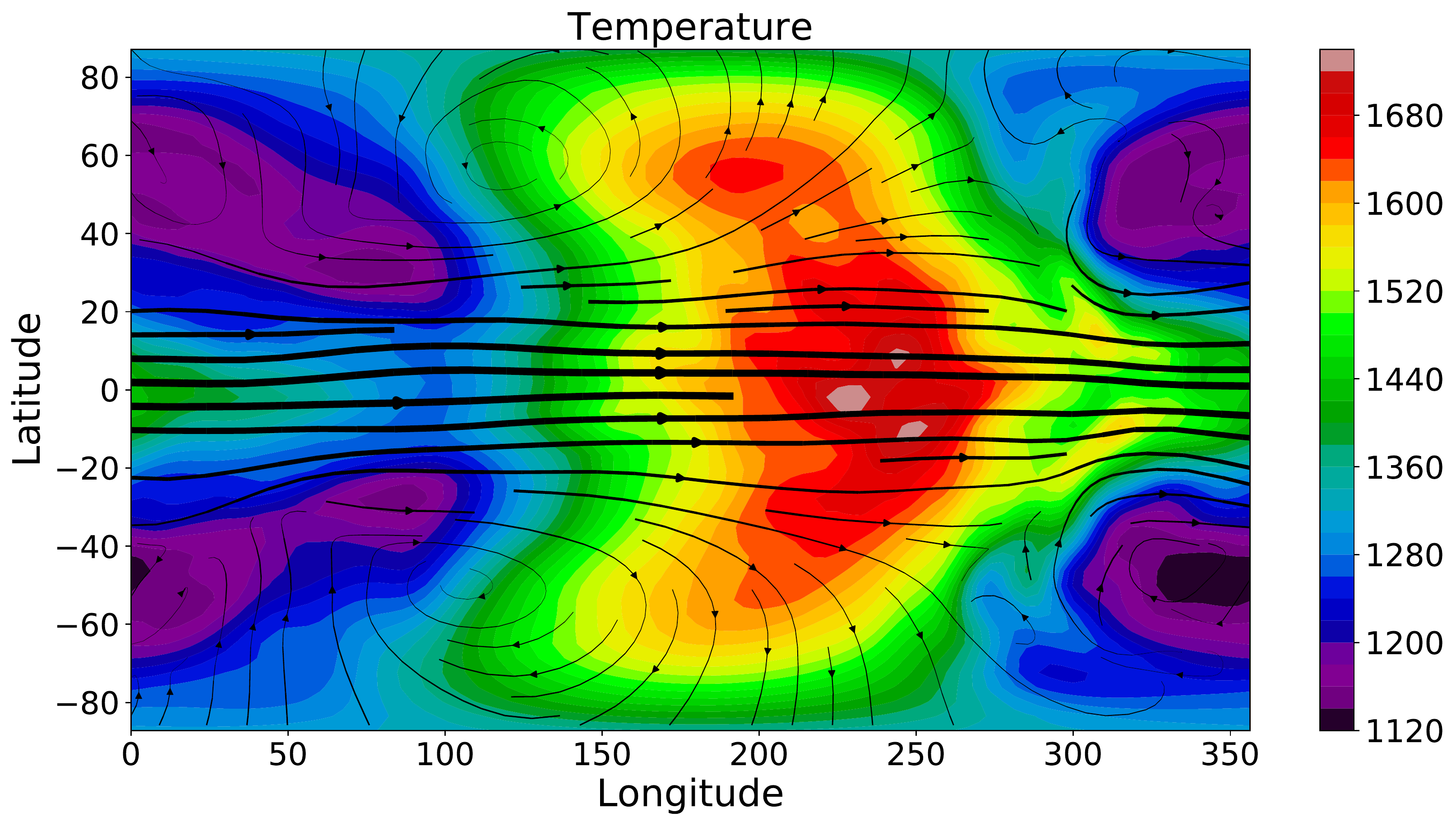}
	\includegraphics[width=\columnwidth]{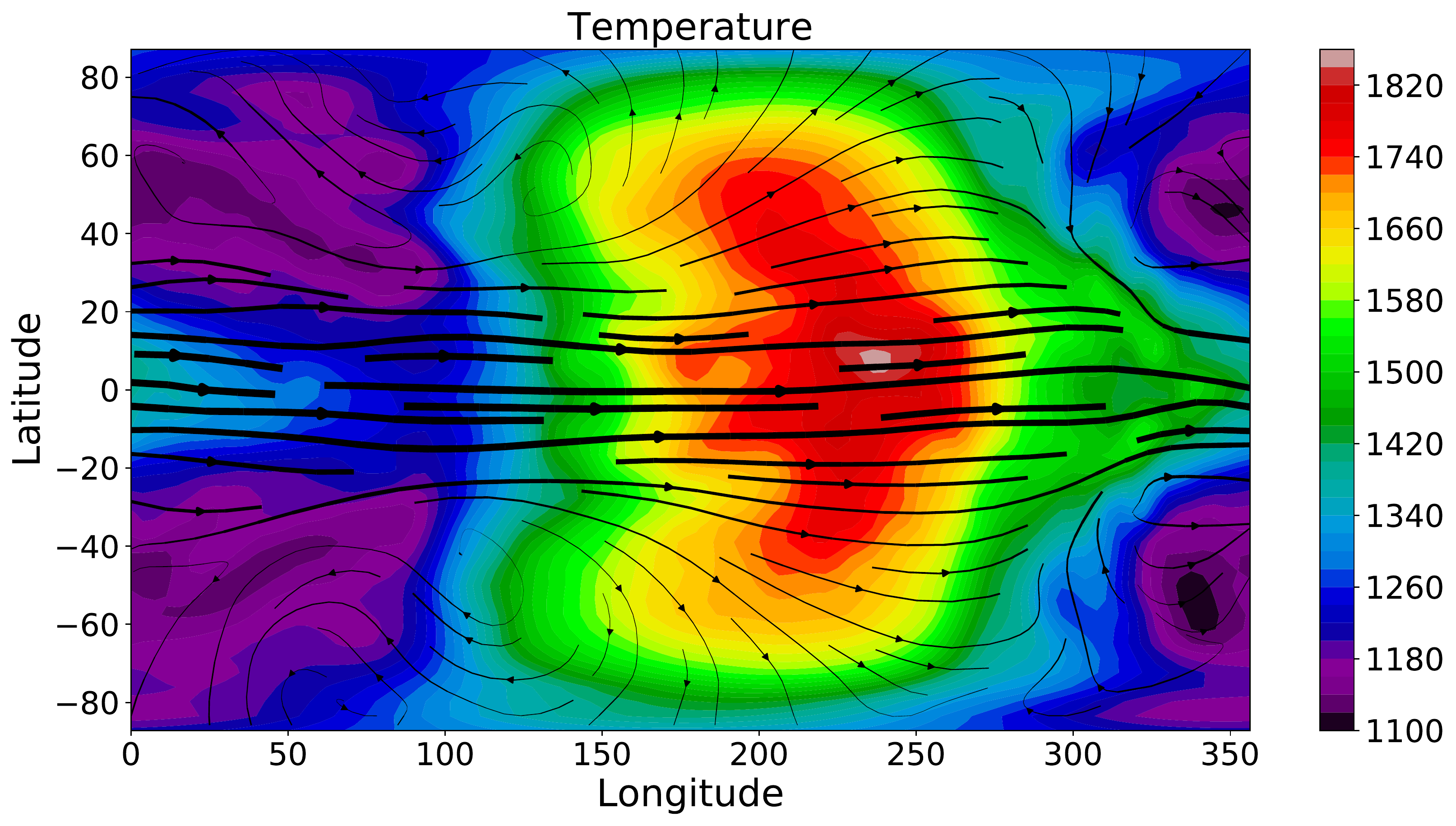}
    \caption{Temperature map (in K) at the 260mb pressure level at planet day 600 in our T31L30 models of HD189733b (top panel), HD209458b (middle) and Kepler7b (bottom).  Superimposed flow lines have a thickness scaled to the wind speed.  The cooler HD189733b experiences stronger eastward heat advection than the hotter HD209458b and Kepler7b.}
    \label{fig:zero2}
\end{figure}

\begin{figure}
	%\includegraphics[width=\columnwidth]{UZ_profile_noVD.pdf}
	\includegraphics[width=\columnwidth]{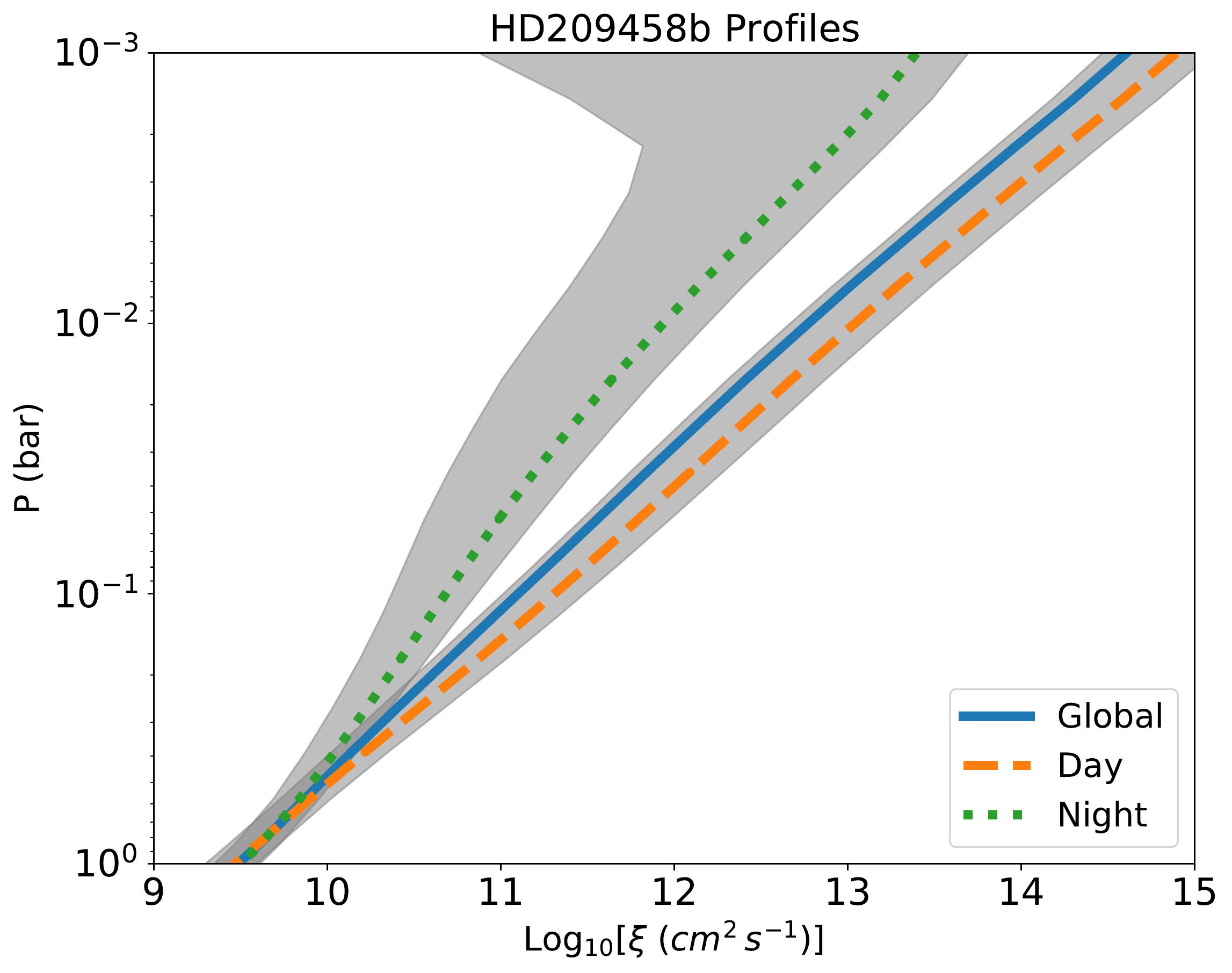}
    \caption{Vertical profiles of thermal diffusivity, $\xi$, in our HD209458b model.  Day-side averaged and night-side averaged profiles are shown as dashed and dotted lines,  respectively, with their corresponding standard deviation (shaded areas). Thermal diffusivity depends sensitively on atmospheric temperatures.  Values above $\sim 0.01$-$0.1$~bar are not necessarily useful as the atmosphere becomes transparent to its own thermal radiation. }
    \label{fig:zero3}
\end{figure}

\begin{figure}
	%\includegraphics[width=\columnwidth]{UZ_profile_noVD.pdf}
	\includegraphics[width=\columnwidth]{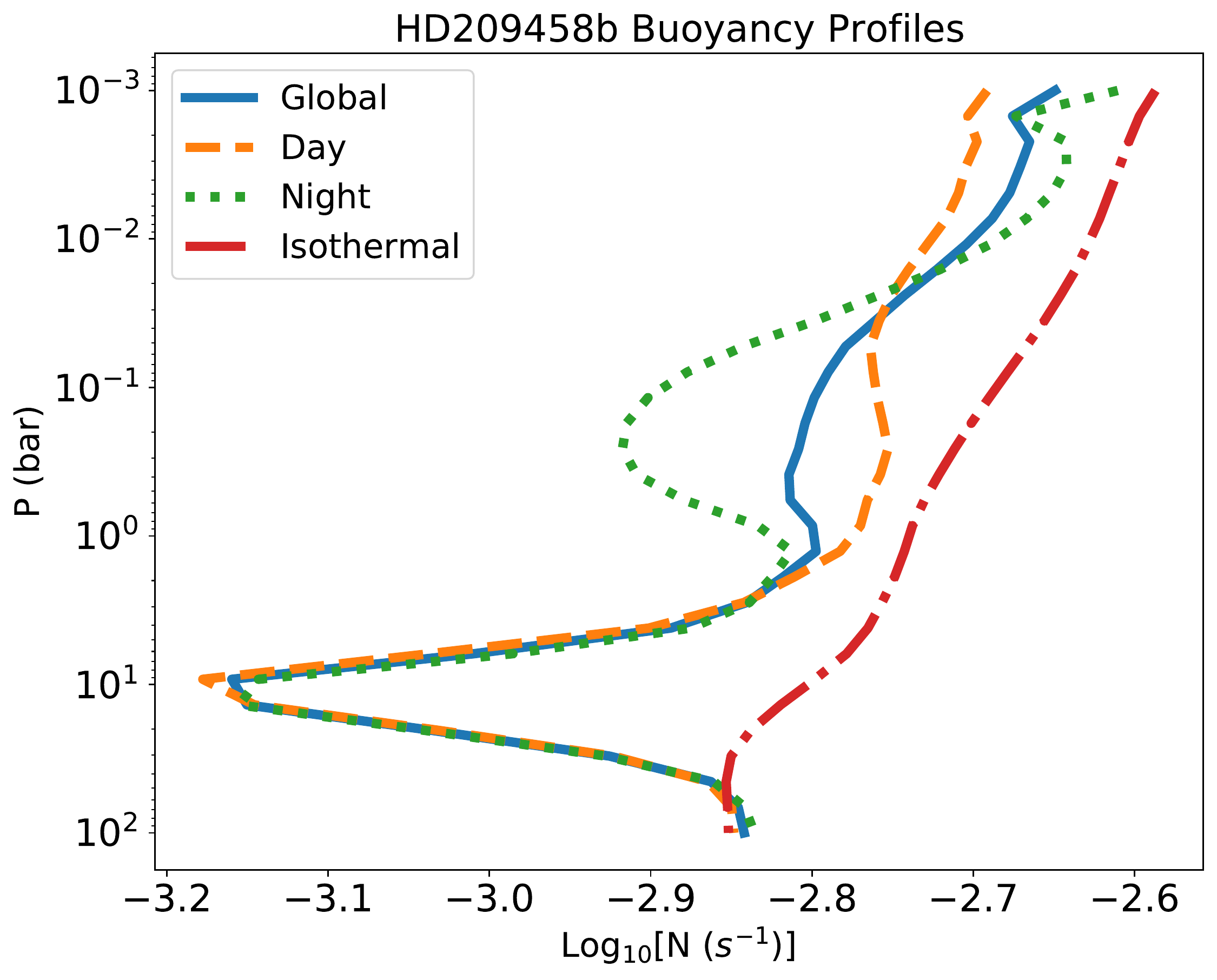}
    \caption{Vertical profiles of buoyancy (Brunt-Vaisala) frequency, $N$,  in our HD209458b model.  Detailed profiles for the day-side only (dashed line), the night-side only (dotted), and a global average (solid),  are compared to an idealized profile that assumes vertical isothermality (dash-dotted).  The horizontal axis displays a very narrow range of values,  highlighting features of the profiles that are otherwise fairly similar.  At the atmospheric top,  night-side stratification is stronger than on the dayside, from smaller pressure scale-heights. In the $0.1$-$1$~bar region, strong heat advection reverses this trend. Profiles below a few bars are not reliable because the deeper atmospheric regions are not relaxed.  A simple isothermal assumption provides an acceptable approximation for our analysis.}
    \label{fig:zero4}
\end{figure}

\begin{figure}
	%\includegraphics[width=\columnwidth]{UZ_profile_noVD.pdf}
	\includegraphics[width=\columnwidth]{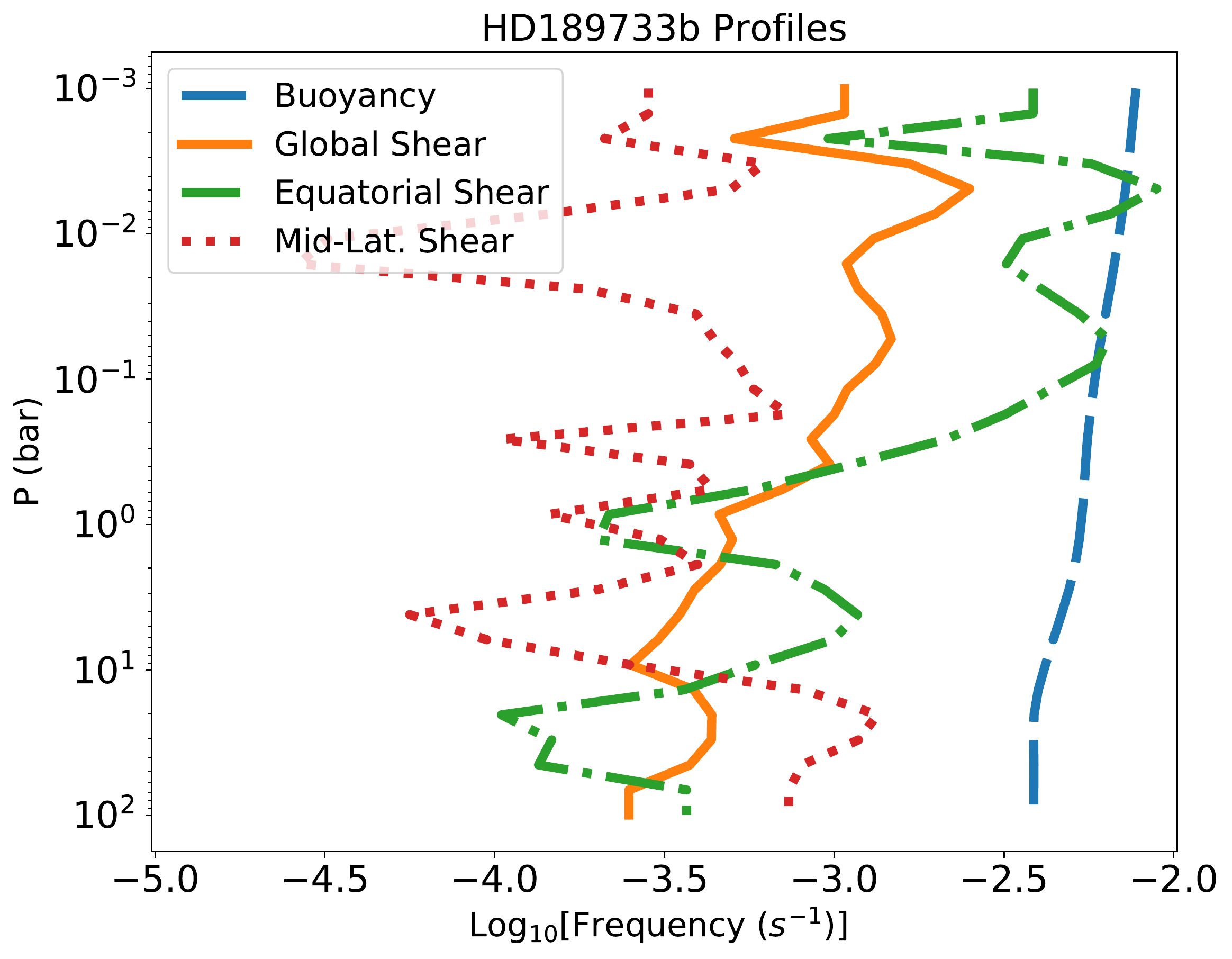}
	\includegraphics[width=\columnwidth]{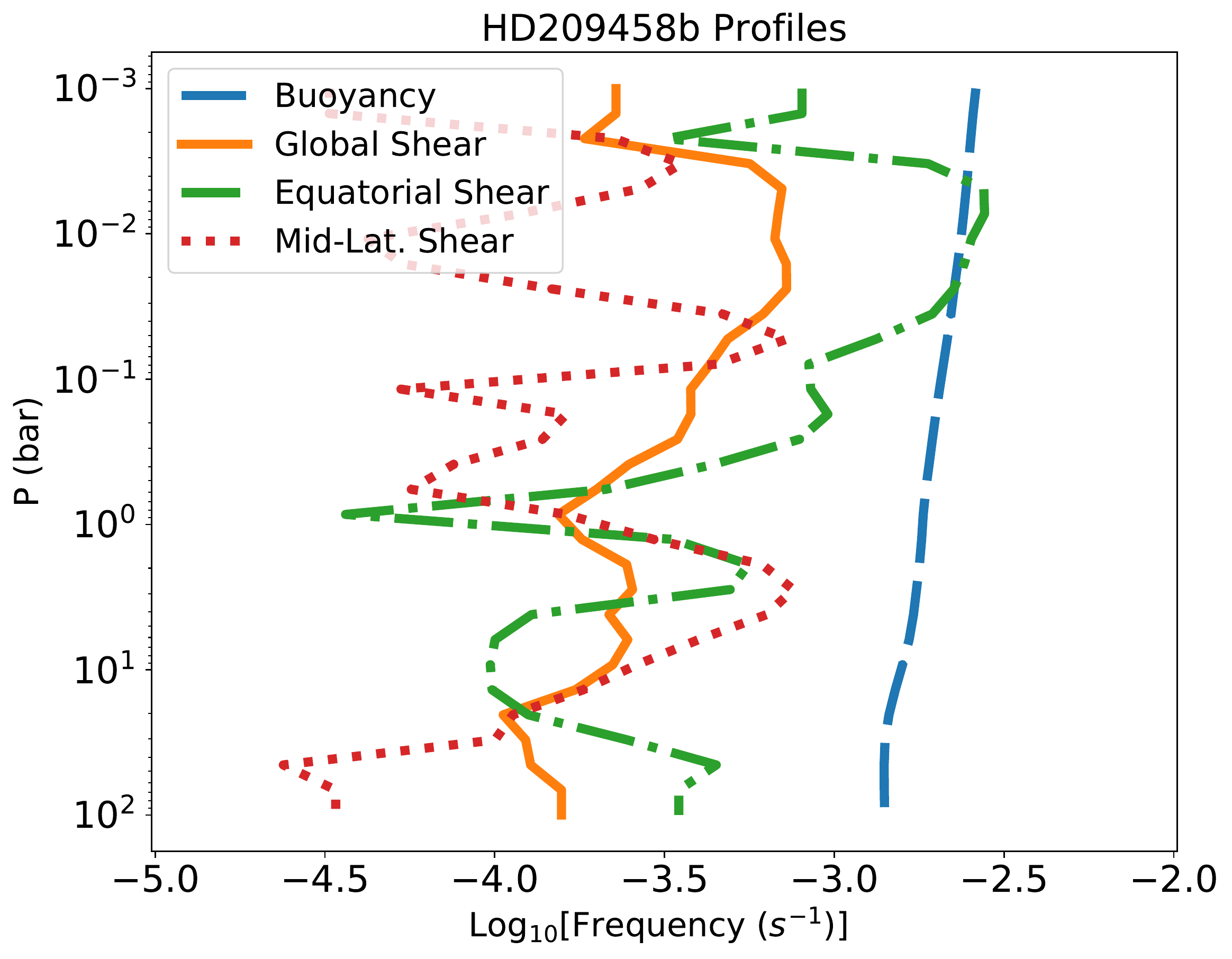}
	\includegraphics[width=\columnwidth]{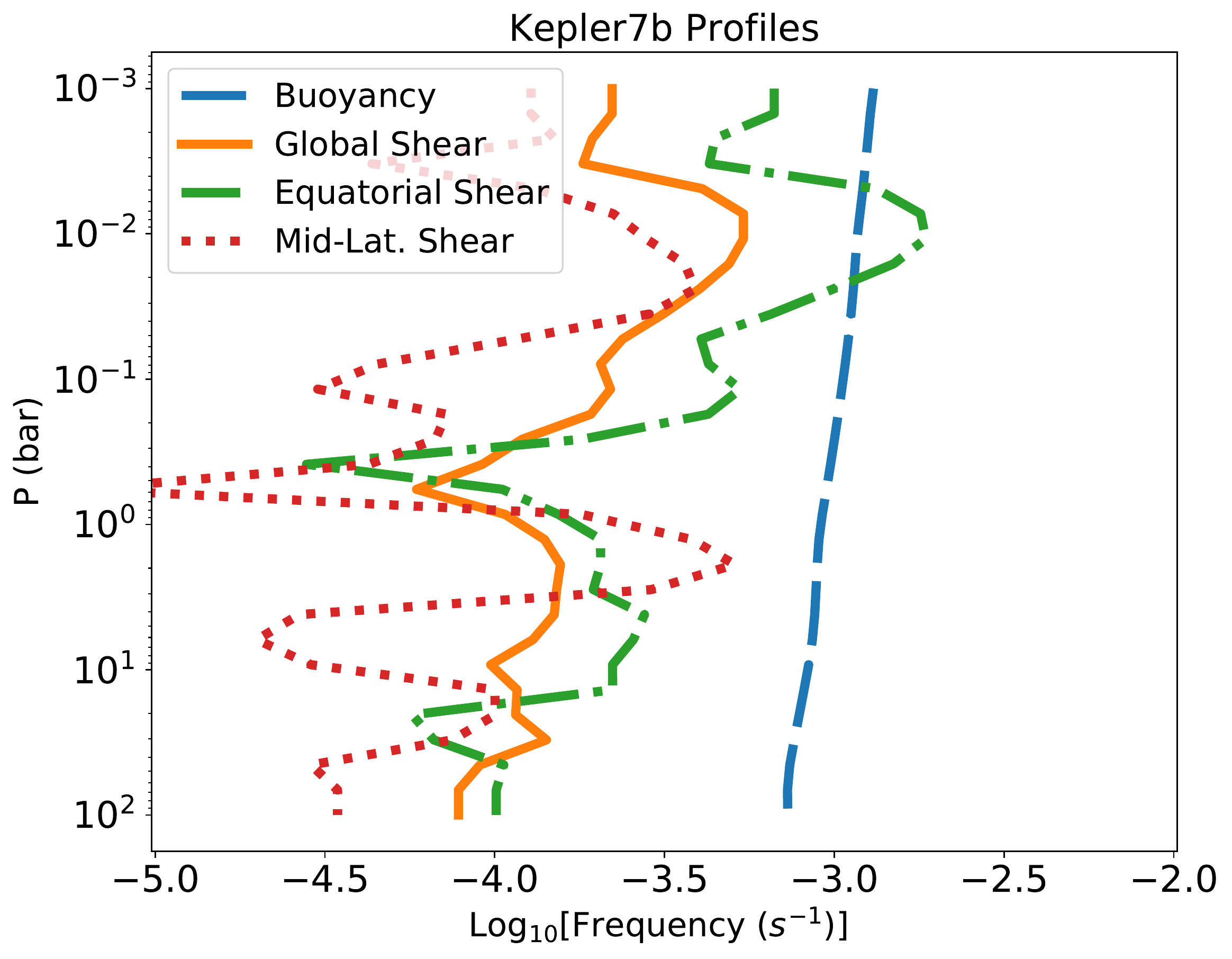}
    \caption{Vertical profiles of buoyancy frequency, $N$,  and shear frequency,  $S$,  in our models of HD189733b (top panel), HD209458b (middle) and Kepler7b (bottom).  In each panel, a unique isothermal buoyancy profile is shown as a long-dashed line, with three zonal wind shear profiles:  at the equator (dash-dotted), at mid-latitudes (dotted) and a globally-averaged version (solid). Equatorial shear is typically a factor 3-10 times stronger than at mid-latitudes. Richardson numbers for the atmospheric mean flow have values Ri$\sim 10-100$. Shear profiles in the top 1-1.5 dex pressure range are affected by upper boundary effects (see Appendix~B for details).}
    \label{fig:zero5}
\end{figure}

To help ground our discussion in more specific estimates, we frame all our results in the context of three hot Jupiter global circulation models for  HD209458b, HD189733b and Kepler7b. Our focus on these three specific planets was dictated by (i) their status as fairly well-characterized cases in the overall population of known Hot Jupiters and (ii) their usefulness in allowing us to sample a range of gravitational accelerations and atmospheric temperatures,  two key parameters affecting the range of vertical transport and mixing expected.  As seen in Table~1, HD189733b,  HD209458b,  and Kepler7b follow a useful hierarchy of equilibrium temperatures, with $T_{\rm eq} \simeq 1200,\, 1450$ and $1630$K respectively \citep{2018haex.bookE.116P}, while sampling a range of gravitational acceleration.

Our global circulation modeling builds on \cite{2020MNRAS.493.5038M}, where a moderate resolution (T31L31 ) model of HD209458b using the Plasim-Gen climate model was described. PlaSim-Gen is an adaptation of the PlaSim Earth-centric GCM to the modelisation of generic gaseous giant planets such as hot Jupiters. PlaSim-Gen was described in \cite{2020MNRAS.493.5038M} and, unless otherwise specified, all model configuration and default parameters are identical to  those discussed in that work.

Here, we add two models, at similar T31L31 resolution, for HD189733b and Kepler7b. As was done for the HD209458b model, we parametrize the radiative forcing of the atmosphere with a Newtonion relaxation scheme, with parameters adjusted to match the leading-order thermal response expected for HD189733b and Kepler7b.  Like in \cite{2020MNRAS.493.5038M}, we use pressure-dependent polynomial fits to $T_{\rm day}$,  $ T_{\rm night}$ and $\tau_{\rm rad}$ for HD209458b from \cite{2011MNRAS.413.2380H}. We rescale the relaxation temperature and relaxation times for HD189733b and Kepler7b in proportion to the surface gravity and typical temperature of the two new planets,  according to
\begin{eqnarray}
T_{\rm day}, \,  T_{\rm night} & \propto & T_{\rm eq} \\
\tau_{\rm rad} & \propto & \frac{P}{g} \frac{C_p}{4 \sigma T^3} \propto \frac{1}{g T_{\rm eq}^3},
\end{eqnarray}
where $P$ is the pressure, $g$ is the gravitational acceleration, $C_p$ is the heat capacity at constant pressure and $\sigma$ the Stefan-Boltzman constant.  This scaling accounts for different absorber columns at a given pressure ($\propto P/g$) and the leading order temperature dependencies. The net result of this scaling are relaxation temperatures scaled by x 0.82 (HD189733b) and x 1.11 (Kepler7b) and radiative relaxation times scaled by x 0.67 ((HD189733b) and x 1.42 (Kepler7b). These scalings assume for simplicity that thermal grey opacities are the same for all three planetary atmospheres of interest.  The strongly idealized radiative forcing that characterizes our models yield reasonable atmospheric forcing conditions for HD189733b and Kepler7b though it is clearly no replacement for more detailed radiative transfer calculations.  

Figures~1 and 2 show zonal wind profiles and temperature-flow maps for our two new hot Jupiter models,  together with that for HD209458b.  Our new models compare reasonably well to previously published models for HD189733b and Kepler7b \citep[e.g.,][]{2009ApJ...699..564S,2017ApJ...850...17R}.  While more advanced models with detailed radiative transfer solutions would improve upon the results shown in Figs~1 and 2, we believe that our models capture enough salient features of the atmospheric flow for us to investigate the general vertical mixing process in all three planets of interest.

To separate physical effects and isolate their contributions,  we note that our models do not account for any magnetic/Lorentz drag on the atmospheric flows \citep{2010ApJ...719.1421P,2013ApJ...764..103R}. Magnetic drag effects could range from negligible for the cooler HD189733b to substantial for the hotter Kepler7b \citep{2012ApJ...745..138M}.

\subsection{Atmospheric Flow Diagnostics}

We have found that some of the diagnostic quantities of interest for our analysis can be challenging to compute reliably,  given the fluctuating/time-dependent nature of the atmospheric flow in our models.  As a result,  we have relied on averaging for some of the quantities of interest.  Throughout our analysis, vertical profiles are obtained by averaging over model longitudes and latitudes.  In some cases,  dayside-only and nightside-only averages are computed as well,  to highlight the very different physical regimes that result from synchronous tidal locking on hot Jupiters.   Some of the averaged vertical profiles are displayed with an associated one standard deviation range,  shown as a shaded region.

Figure~3 shows vertical profiles of the thermal diffusivity in our HD209458b model:
\begin{equation}
\xi \equiv \frac{16}{3} \frac{\gamma -1}{\gamma} \frac{\sigma T^4}{\kappa \rho P},
\end{equation}
where $\kappa$ is the radiative opacity.
Day and night values at a given pressure level can differ by 1-3 orders of magnitude,  owing to the strong temperature dependence.  Note that values higher up than the 0.1-0.01 bar level should be used with caution because the atmosphere becomes transparent to its own radiation there.

Gradients quantities can be noisy, from the time-variable nature of the atmospheric flow and from the numerical differentiation process.  We have found that a reliable proxy for the thermal buoyancy profiles $N$ in our models, defined by 
\begin{equation}
N^2 = \frac{g}{\gamma} \frac{d}{dz} \ln \frac{P}{\rho ^ \gamma}
\end{equation}
is to compute the Brunt-Vaisala frequency for the equivalent isothermal atmosphere:
\begin{equation}
N_{\rm iso}^2 = \frac{\gamma -1}{\gamma} \frac{g}{H_p},
\end{equation}
where we adopt the diatomic value $\gamma =7/5$ for the adiabatic index and the pressure scale-height is defined as
\begin{equation}
H_p = \frac{R_{\rm gas} T} {g}.
\end{equation}

The shear rate $S$ is computed as the vertical gradient of the zonally-averaged profile of the zonal wind $U$ only (i.e.,  zonal component of the wind vector,  averaged over longitudes):
\begin{equation}
S = \frac{dU}{dz}.
\end{equation}

This implies that day-night variations are averaged over when computing the vertical shear in our analysis (following Eq.~7)..  Other diagnostic quantities of interest are not zonally-averaged,  unless otherwise specified,  and thus retain information about the day-night variations.

Figure~4 shows detailed buoyancy profiles in our model of HD209458b. The day-side,  night-side and globally-averaged profiles are consistent with each other and close to the idealized isothermal-equivalent profile mentioned above.  Thermal buoyancy increases slightly with height in the atmosphere.  In what follows, we only use the isothermal-equivalent profile,  for simplicity.

Figure 5 compares the buoyancy profile to the vertical shear profiles of the atmospheric flow in our models of HD209458b,  HD189733b and Kepler7b. To illustrate the stronger shear typically present in the equatorial regions of our models,  we show shear profiles at the equator ($0$~deg latitude), at mid-latitudes ($45$~deg latitude) and a globally-averaged profile.  The profiles of equatorial shear are often above the average atmospheric shear profiles (though not systematically),  which indicates that vertical mixing from shear instabilities may preferentially develop and have a stronger impact in the equatorial regions.

In the models for all three planets,  we find Richardson numbers 
\begin{equation}
Ri \equiv \frac{N^2}{S^2} \sim 30$-$50
\end{equation}
in the bulk of the atmospheric flow.  Richardson number values well above unity imply that the flow is stable against dynamical shear instabilities. Provided sufficiently fast heat transport exists in the medium to neutralize the strongly stabilizing stratification ($N \gg S$), the possibility for secular shear instabilities to develop remains, which is the main topic of interest in this work. 

The systematically stronger shear at the top of all three models is suspicious and we have confirmed with a vertically extended model described in Appendix~B that the upper level shear is sensitive to the location of the model top (i.e., the proximity to the upper boundary). Results in the bulk of the atmospheric domain remain unaffected, however, and we focus on those in our analysis.

\section{Dominant Uncertainties for Secular Shear Transport} 

Double-diffusive shear instabilities, also known as secular shear instabilities in the stellar astrophysics literature, are one of the main candidate mechanisms for angular momentum transport in stars \citep[e.g.,][and references therein]{2013ApJS..208....4P,2000ApJ...528..368H}. As emphasized by \cite{2019MNRAS.485L..98M} -- see also \cite{2010ApJ...725.1146L} -- in the exoplanetary atmospheric context, the key physical ingredients making these instabilities relevant are (i) a strong enough vertical shear relative to the thermal stratification and (ii) a strong thermal diffusivity enabled by photon radiative transport in the hot and optically thick atmospheric gas.

There are are at least two differences  between the exoplanet atmosphere and stellar cases that could impact the applicability of existing stellar-specific results to the exoplanet context. First, while shear is typically mechanically imposed in the stellar evolution context (from an external torque), it is typically thermally driven in the form of differential wind speeds in the hot Jupiter context.\footnote{Magnetic torques from Lorentz forces in sufficiently hot and ionized atmospheres could provide further analogy with the stellar context \citep{2010ApJ...719.1421P,2013ApJ...764..103R,2012ApJ...745..138M}.} Second, the optically-thick condition assumed in the double-diffusive shear formulation of stellar astrophysics is bound to break down as one approaches the thermal photosphere of an exoplanet \citep{2019MNRAS.485L..98M}.

It is unclear whether the manner by which a given shear profile is imposed (whether mechanically or thermally) will influence the resulting secular shear regime. Consistency in the results of direct numerical simulations of secular shear turbulence with bulk forcing \citep{2013A&A...551L...3P} and boundary forcing \citep{2017ApJ...837..133G} suggests that the answer may be no. By contrast, the different thermal behaviour of a transparent gas with optically-thin turbulent fluctuations, relative to an opaque gas with optically-thick fluctuations, could have a major effect on the resulting secular shear turbulence, since its existence relies on efficient damping of thermal perturbations \citep{1958JFM.....4..361T, 1974JFM....64...65D, 1974IAUS...59..185Z}. We address this issue in detail below.

\subsection{Semi-Transparent Regime of Shear Turbulence}

\subsubsection{Relevant Lengthscales}
\label{sec:length}

In the vicinity of the thermal photosphere (and above), one expects any turbulent fluctuation to become optically-thin, even for fluctuations approaching the size of the local pressure scale height, $H_p$.  Even below the photosphere, however, in atmospheric regions that are formally opaque, a turbulent fluctuation of small enough scale can be optically thin to its own thermal radiation \citep[e.g.,][]{1957ApJ...126..202S}.\footnote{While some of our reasoning technically rests on linear, infinitesimal perturbations, we note that even in the presence of strong finite-amplitude density fluctuations, say $\delta \rho / \rho \sim 0.5$,  our conclusions would largely hold.  Also, secular shear is expected to operate in a Boussinesq-like, quasi-incompressible regime\citep{1960ApJ...131..442S,2013A&A...551L...3P,2017ApJ...837..133G}} 

We use the photon mean free path as the limiting scale for transparent radiative transport:
\begin{equation}
\lambda_{\rm ph} = \frac{1}{\rho \kappa_P},
\end{equation}  
where $\rho$ is the local mass density and $\kappa_P$ is the Planck-mean grey opacity. Any turbulent fluctuation with a characteristic scale $l < \lambda_{\rm ph}$ is expected to behave thermally in an optically-thin manner. We define a complementary limiting scale for opaque radiative transport:
\begin{equation}
\lambda_{\rm Ross} = \frac{1}{\rho \kappa_R},
\end{equation}  
 where $\kappa_R$ is the Rosseland-mean grey opacity, so that a turbulent fluctuation with a characteristic scale $l > \lambda_{\rm Ross}$ is expected to behave thermally in an optically-thick (diffusive) manner.
 
 A large range of turbulent scales might generally be expected to be present in a secular-shear turbulent medium with high Reynolds number, from the largest turbulence integral scale down to the turbulence (viscous) dissipation scale \citep[][see also Appendix D]{1987flme.book.....L}. The nature of thermal transport in the turbulence thus depends on a range of scales up to the integral scale. The integral scale may be related to large unstable modes feeding kinetic energy to the turbulent cascade, or may be related to a globally limiting scale, such as the pressure scale height,   $H_p$,  when the underlying instability is local in nature.  Lacking an a priori understanding of secular shear turbulence, we will start by considering a few scales that are potentially relevant as turbulence integral scales.  
  
The first scale of interest, that we refer to as the buoyancy scale, is the scale at which the thermal diffusion time of an optically-thick perturbation, $\sim l^2/\xi$ for thermal diffusivity $\xi$, equals the thermal buoyancy restoring time, $N^{-1}$ (the inverse of the Brunt-Vaisala frequency $N$). The resulting buoyancy scale is 
 \begin{equation}
\lambda_{\rm buoy} = \sqrt{\frac{\xi}{N}}.
\end{equation}  
For large-enough, and thus opaque, perturbations in a semi-transparent medium\footnote{We refer to gas as being semi-transparent when large scale perturbations behave radiatively as opaque while small scale perturbations behave such that they are transparent to their own thermal radiation \citep{1957ApJ...126..202S}} this scale is expected to be about the largest one beyond which perturbations are not able to thermally diffuse fast enough that they can neutralize the stabilizing role of thermal buoyancy. 
 
Another potentially relevant scale can be similarly derived from the shear rate, $S$, which is a proxy for the growth rate of a shear-based instability in the absence of buoyancy. We obtain the shear scale as
 \begin{equation}
\lambda_{\rm shear} = \sqrt{\frac{\xi}{S}}.
\end{equation}  
 
Figure~\ref{fig:length} shows vertical profiles for the four relevant scales we just introduced, computed from our atmospheric models of HD209458b, HD189733b and Kepler7b.  When computing the transparent and opaque radiative scales $\lambda_{\rm ph}$ and $\lambda_{\rm Ross}$,  a horizontal average of $\rho$ was used for simplicity. A Planck-mean opacity $\kappa_P = 0.5$~cm$^2$~g$^{-1}$ was used throughout, which represents fairly well the typical values of $\sim 0.5-0.9$cm$^2$~g$^{-1}$ computed by \cite{2014ApJS..214...25F} for the range of pressure and temperature $1$-$100$mbar and $1000$-$2000$K. A Rosseland-mean opacity $\kappa_R = 10^{-2}$cm$^2$~g$^{-1}$ was used throughout, for consistency with the typical grey opacity used in 1D studies and GCMs of HD209458b \cite[e.g.][]{2010A&A...520A..27G,2012ApJ...750...96R}. Our conclusion are not expected to be significantly affected by using more detailed opacity data. The region between $\lambda_{\rm ph}$ and $\lambda_{\rm Ross}$ has been shaded, to highlight the transition from optically-thin to optically-thick regimes.

Figure~\ref{fig:length} shows that near photospheric regions, over the $10^{-3}$-$10$~bar pressure range, the buoyancy and shear scales exceed the atmospheric pressure scale height, $H_p$, for all three atmospheres of interest. As a result, the turbulence integral scale might end up being limited by global effects.  For concreteness, we use a limiting scale of  $0.3 H_p$ in our analysis,  shown as a vertical dotted line in each of the three panels of Figure~\ref{fig:length}.

Small-scale turbulent fluctuations, found to the left of the slanted shaded area, will behave radiatively in an optically-thin manner, while larger scales to right of the shaded region  will behave in an optically-thick manner. We conclude that secular shear turbulence will generally be in a semi-transparent regime for the three hot Jupiter atmospheres considered here. High-up in these atmospheres ($P \sim 1$mbar), above the thermal photospheres located around $\sim 100$mbar (HD189733b) to $\sim 10$mbar (Kepler7b), all turbulent fluctuations up to scales $\leq H_{\rm p}$ are optically-thin. Deeper in these atmospheres, a mix of marginally optically-thick to semi-transparent regimes are possible, potentially combining opaque integral scales together with transparent turbulent fluctuations on smaller scales.

Note that it is possible that the turbulence integral scale for secular shear instability is significantly smaller than the limiting global scale (i.e $\ll H_{\rm p}$), which would make the optically-thin regime even more relevant to the secular shear turbulence regime expected in hot Jupiter atmospheres.

To summarize, turbulent perturbations are likely to be optically-thin or in a mixed optically-thin/thick (= semi-transparent) regime for much of the Hot Jupiter atmospheric parameter space. This calls for a revision of our results in \cite{2019MNRAS.485L..98M} to better account for the semi-transparent regime.

\subsubsection{Relevant Timescales}

Let us now turn to a discussion of various timescales thought to be relevant to secular shear instabilities and turbulence.

For secular shear perturbations to grow, one expects the need for a perturbed fluid element to be able to neutralize its stabilizing buoyancy, through fast-enough thermal exchange with its environment. This is achieved through scale-dependent heating/cooling\footnote{The thermal diffusion time scales as the square of the effective size of the perturbed fluid element,  $\sim l^2 / \xi$, as is the case in classic diffusion problems. } for optically-thick scales satisfying the radiative diffusion approximation, and through scale-independent heating/cooling for optically-thin scales \citep{1957ApJ...126..202S,1974JFM....64...65D}.

Linear stability analysis of secular shear instability has proven challenging, as compared to other double-diffusive instabilities \citep{1967ApJ...150..571G,1968ZA.....68..317F,2013ApJ...768...34B}. We are guided in our discussion by results obtained by \cite{1974JFM....64...65D} for the semi-transparent regime and \cite{2010ApJ...725.1146L} for the optically-thick regime. In particular,  \cite{2010ApJ...725.1146L} show growth rates approaching  a tenth of $\tau_{\rm shear} ^{-1}$ for the most unstable modes in the limited parameter space they were able to numerically survey.

A relevant timescale to compare to the thermal exchange timescale might then be the inverse growth rate of a shear-unstable perturbation,  $\tau_{\rm shear} \sim S^{-1}$, if our focus is on allowing thermal exchanges fast enough for the growth of shear unstable modes to proceed unaltered (irrespective of thermal stratification considerations).  Another  relevant comparison timescale is the restoring time from buoyancy, $\tau_{\rm buoy} \sim N^{-1}$, if one is to require rapid suppression of the stabilizing influence of thermal stratification by thermal exchanges.  Since we are concerned with atmospheric flows having Richardson numbers $Ri > 1$,  we will find unsurprisingly that the buoyancy timescale offers the more stringent constraint of the two. 

A different line of reasoning considers the energetics of turbulent fluctuations in establishing what amounts to a generalized Richardson criterion for the stability of secular shear \citep{1958JFM.....4..361T,1974IAUS...59..185Z}. In particular,  \cite{1972ARA&A..10..261S}, in reference to \cite{1964ApJ...139...48M}, has argued that a generalized stability criterion for secular shear  is 
\begin{equation}
\frac{N^2}{S} \tau_{\rm cool}  = Ri S \tau_{\rm cool}  > 1,
\end{equation}    
where $\tau_{\rm cool}$ is the relevant thermal exchange timescale,  whether in the optically-thick ($\tau_{\rm cool} \sim l^2/\xi$ for a perturbation of size $l$) or optically-thin ($\tau_{\rm cool} =\tau_{\rm thin} $) regime. The shortest, scale-independent heat exchange time available to a turbulent fluctuation is the optically-thin thermal relaxation timescale derived by \cite{1957ApJ...126..202S} \citep[see also][for secular shear specifically]{1974JFM....64...65D}:
\begin{equation}
\tau_{\rm thin}  = \frac{C_\text{v}}{16 \kappa_{\rm P} \sigma T^3},
\end{equation} 
 where $C_\text{v}$ is the gas heat capacity at constant volume, $\kappa_{\rm P}$ is the Planck-mean grey opacity and $\sigma$ is the Stefan-Boltzmann constant. This optically-thin thermal exchange timescale depends sensitively on the atmospheric temperature, $T$.  
 
We note that the fundamental thermal relaxation time $\tau_{\rm thin}$ should not be confused with the Newtonian forcing relaxation time $\tau_{\rm rad}$ discussed in \S~2.1 and in Eq.~2.   $\tau_{\rm rad}$ applies to global atmospheric scales and covers both the optically-thick and optically-thin regimes. Only when the atmosphere becomes globally optically-thin do we expect $\tau_{\rm rad}$ to asymptote to the fundamental $\tau_{\rm thin}$ limit. This limit is only approached at the domain top in the models considered here (at low enough pressures).

Figure~\ref{fig:times}  compares vertical profiles of the various timescales we just introduced for our three hot Jupiter atmospheres of interest. Differences emerge between the three cases, most  notably between the cooler HD189733b and the hotter Kepler7b. In all three cases, day-side thermal exchange times are significantly shorter than night-side thermal exchange times.  

As mentioned above, the shear time offers a weaker constraint than the buoyancy time for the atmospheric flows of interest here.  Comparing the buoyancy time vs the thermal exchange time,  Figure~\ref{fig:times} reveals that the dayside and nightside of Kepler7b, and the dayside of HD209458b, have short enough thermal exchange times ($\tau_{\rm thin}  < \tau_{\rm buoy}$) for transparent secular shear instability to potentially operate, while transparent thermal exchanges are essentially too slow on HD189733b for secular shear turbulence to operate anywhere in the atmosphere.  If one uses instead the more stringent Moore-Spiegel constraint for how fast thermal exchanges need to be for secular shear turbulence to operate, we find that it may be restricted to the hotter dayside of Kepler7b and,  at a marginal level, the hottest regions on the dayside of HD209458b, excluding all other atmospheric regions in our three models of interest.

The picture that emerges from this comparison is one where the leading factor determining the likelihood of transparent shear turbulence operating is the atmospheric temperature.  Variations in shear and buoyancy times across the three planets,  with somewhat shorter timescales on HD189733b and somewhat longer on Kepler7b, have comparatively little effect on this conclusion 
as the strong dependence of the transparent thermal exchange time on temperature dominates.

\subsection{Magnitude of Turbulent Transport}

\subsubsection{Double-Diffusive Regime}

Uncertainties on the magnitude of turbulent transport from secular shear turbulence, when it operates, are substantial, even in the currently better studied optically-thick ("double-diffusive") regime. \cite{1974IAUS...59..185Z,1992A&A...265..115Z} has proposed a local model for the resulting turbulence (with shear within stars in mind), according to which the magnitude of cross-shear turbulent transport scales as 
\begin{equation}
K_{\rm zz}^{\rm Zahn}  =0.08  \frac{S^2}{N^2} \xi = 0.08 \, Ri^{-1} \, \xi.
\end{equation} 
Direct numerical simulations have so far supported Zahn's stability criterion for secular shear turbulence and have been consistent with, though not approaching, the level of turbulent transport expected in Zahn's model \citep[see detailed discussions in][]{2017ApJ...837..133G, 2020ApJ...901..146G}.

It is also useful to bound expectations by considering a 'maximal transport' hypothesis, according to which the maximum level of transport that could be achieved yields \citep{2019MNRAS.485L..98M}:
\begin{equation}
K_{\rm zz}^{\rm Max}  = \frac{S}{N} \xi = Ri^{-1/2} \, \xi.
\end{equation} 
The intuition behind this Max scaling for turbulent transport from secular shear turbulence is grounded in a linear growth argument. An unstable shearing mode with wavelength $\lambda < \lambda_{\rm Max} \simeq \lambda_{\rm buoy} =(  \xi/N )^{1/2}$  might be expected to grow in a regime with an effectively neutralized buoyancy and a growth rate asymptotically approaching the shear rate $S$ \citep[e.g.,][]{2010ApJ...725.1146L}. The magnitude of transport that can result in such a scenario is thus expected to be maximized at $\sim S \lambda_{\rm Max}^2 =  K_{\rm zz}^{\rm Max}$. We use this $K_{\rm zz}^{\rm Max}$ as a likely upper limit on the magnitude of transport that can be achieved by double-diffusive shear turbulence. One limitation of the $K_{\rm zz}^{\rm Max}$ formulation is that the turbulence integral scale is likely $< H_{\rm p}$, as mentioned earlier  in \S\ref{sec:length}, so that  $\lambda_{\rm Max} < H_{\rm p}$ and we must have $K_{\rm zz}^{\rm Max} < S H_{\rm p}^2$.

\begin{figure}
	%\includegraphics[width=\columnwidth]{UZ_profile_noVD.pdf}
	\includegraphics[width=\columnwidth]{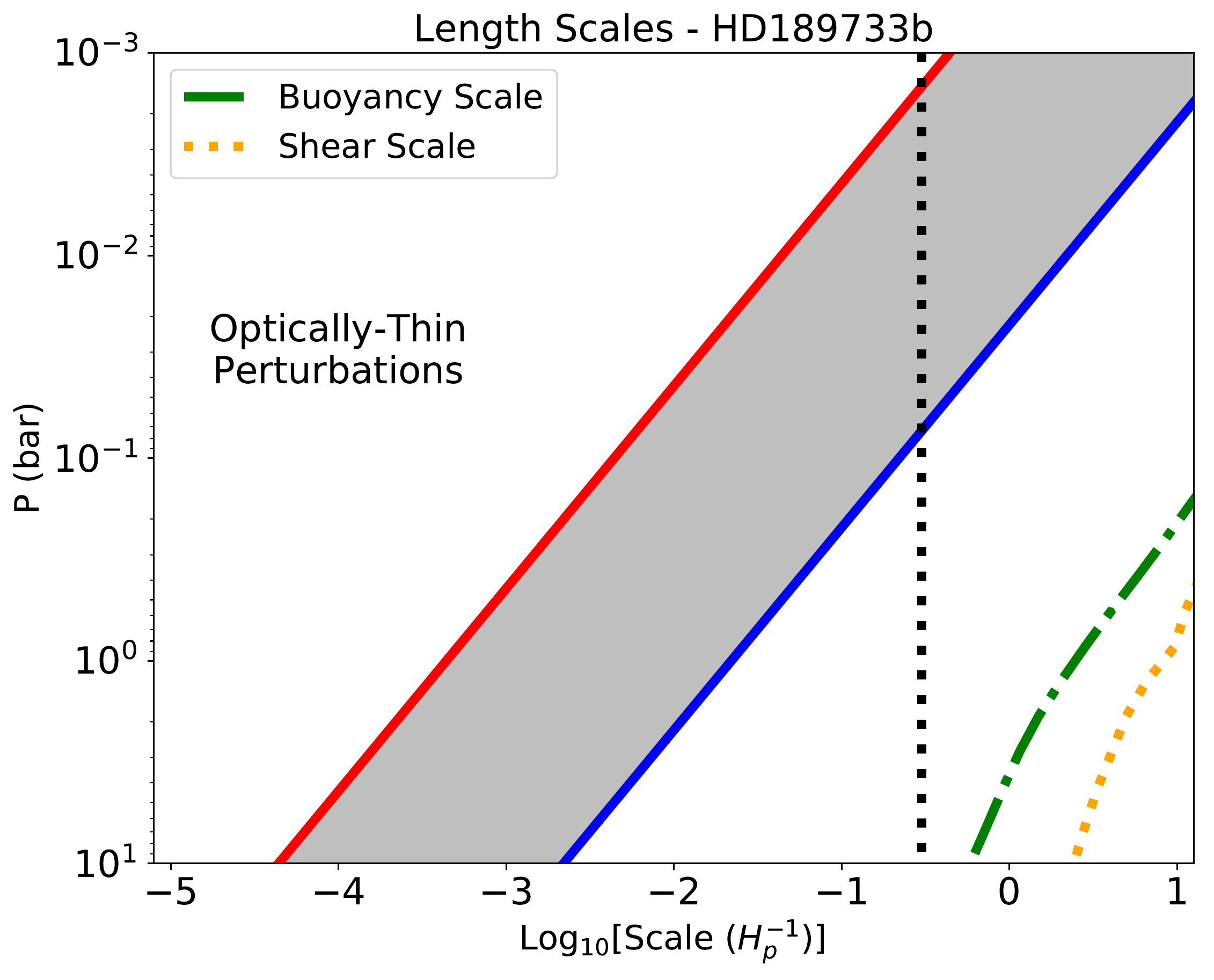}
	\includegraphics[width=\columnwidth]{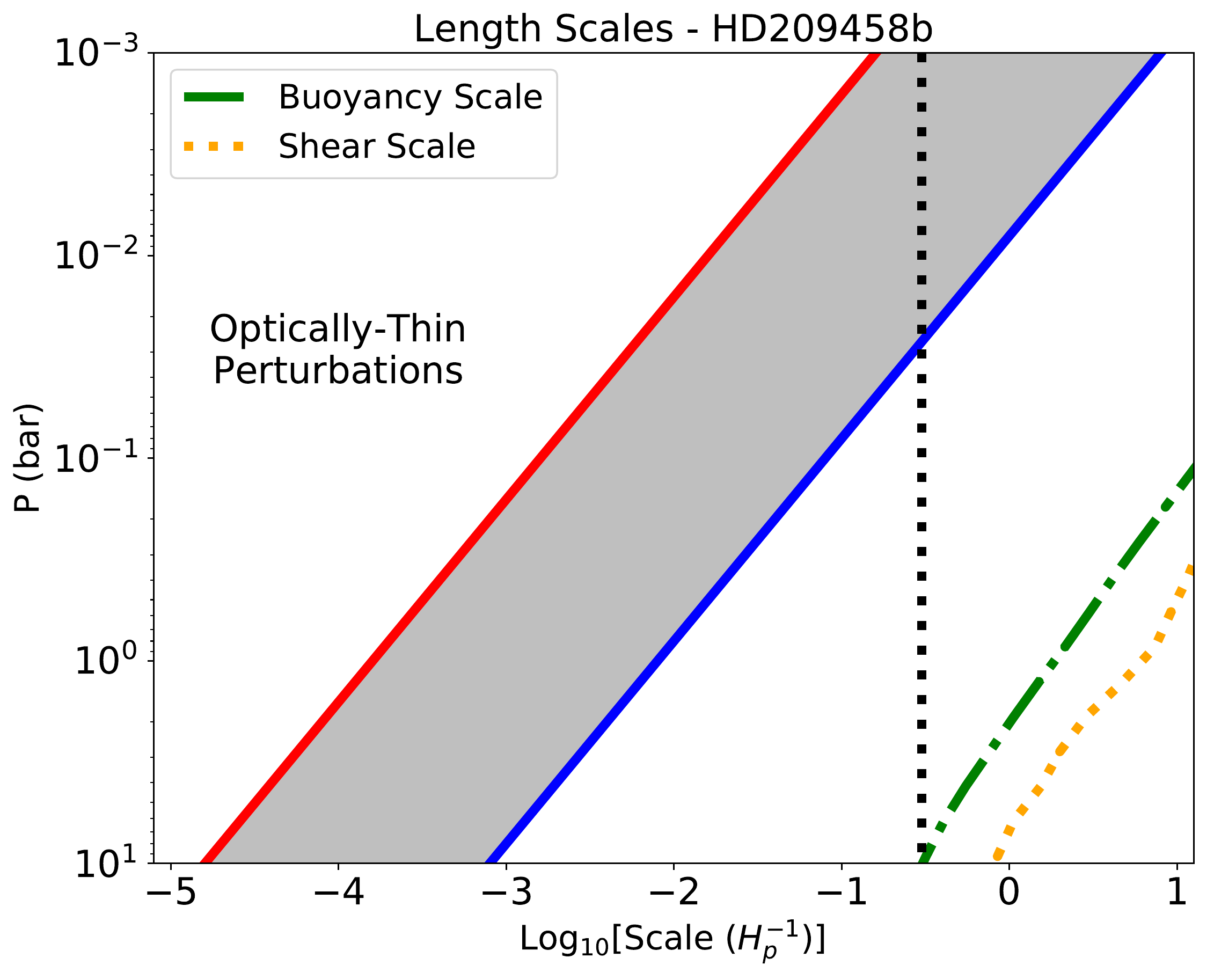}
	\includegraphics[width=\columnwidth]{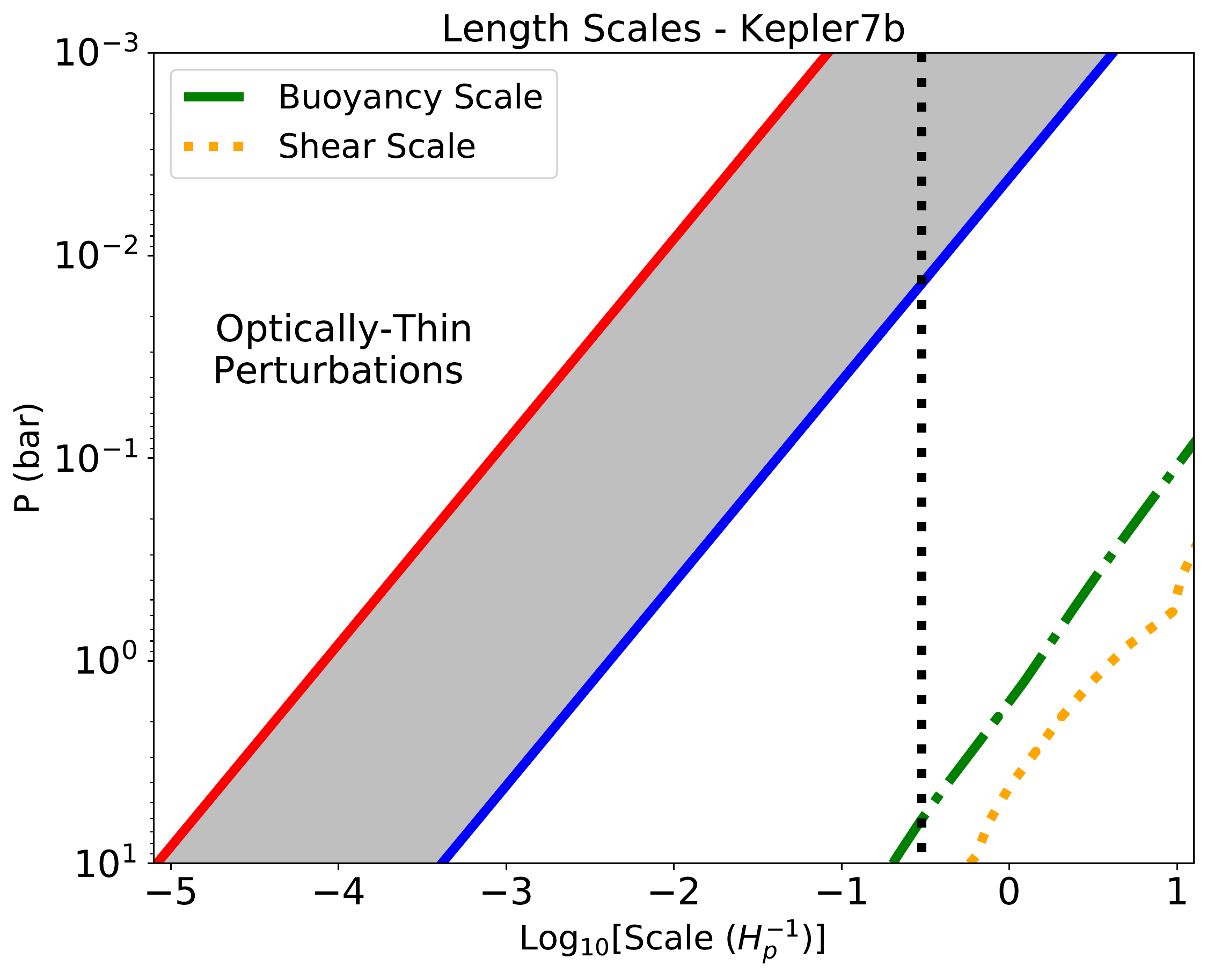}
    \caption{Range of transparent (upper left regions) vs opaque (lower-right regions) perturbation lengthscales in the atmospheres of HD189733b (top panel), HD209458b (middle) and Kepler7b (bottom). The intermediate, semi-transparent regime is highlighted by the shaded region. Vertical profiles of the buoyancy and shear scales, defined in the text, are also shown for comparison. All scales are shown in units of the local pressure scale height, $H_{\rm p}$. A vertical dotted line highlights the limiting value $0.3 H_{\rm p}$.}
    \label{fig:length}
\end{figure}

\begin{figure}
	%\includegraphics[width=\columnwidth]{UZ_profile_noVD.pdf}
	\includegraphics[width=\columnwidth]{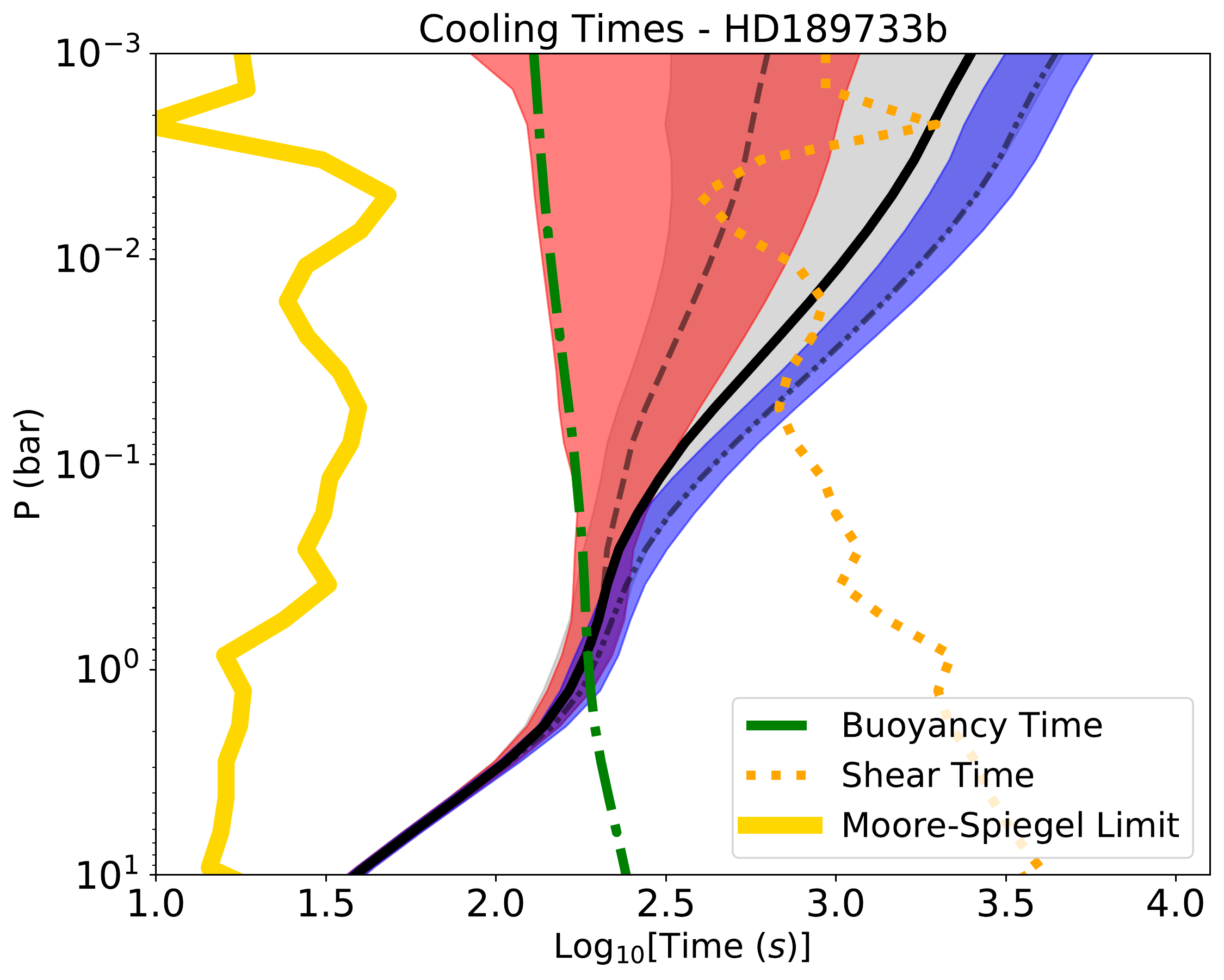}
	\includegraphics[width=\columnwidth]{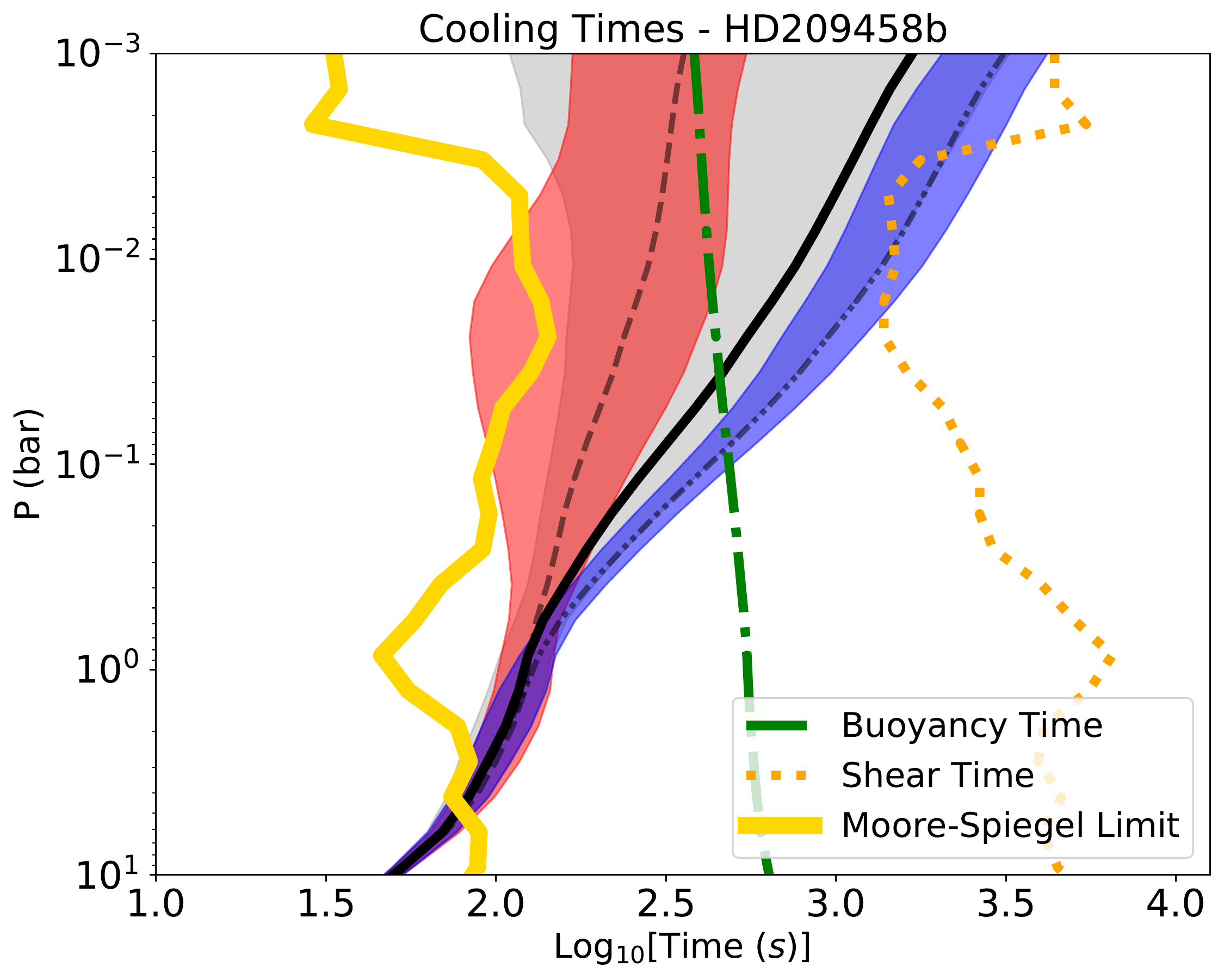}
	\includegraphics[width=\columnwidth]{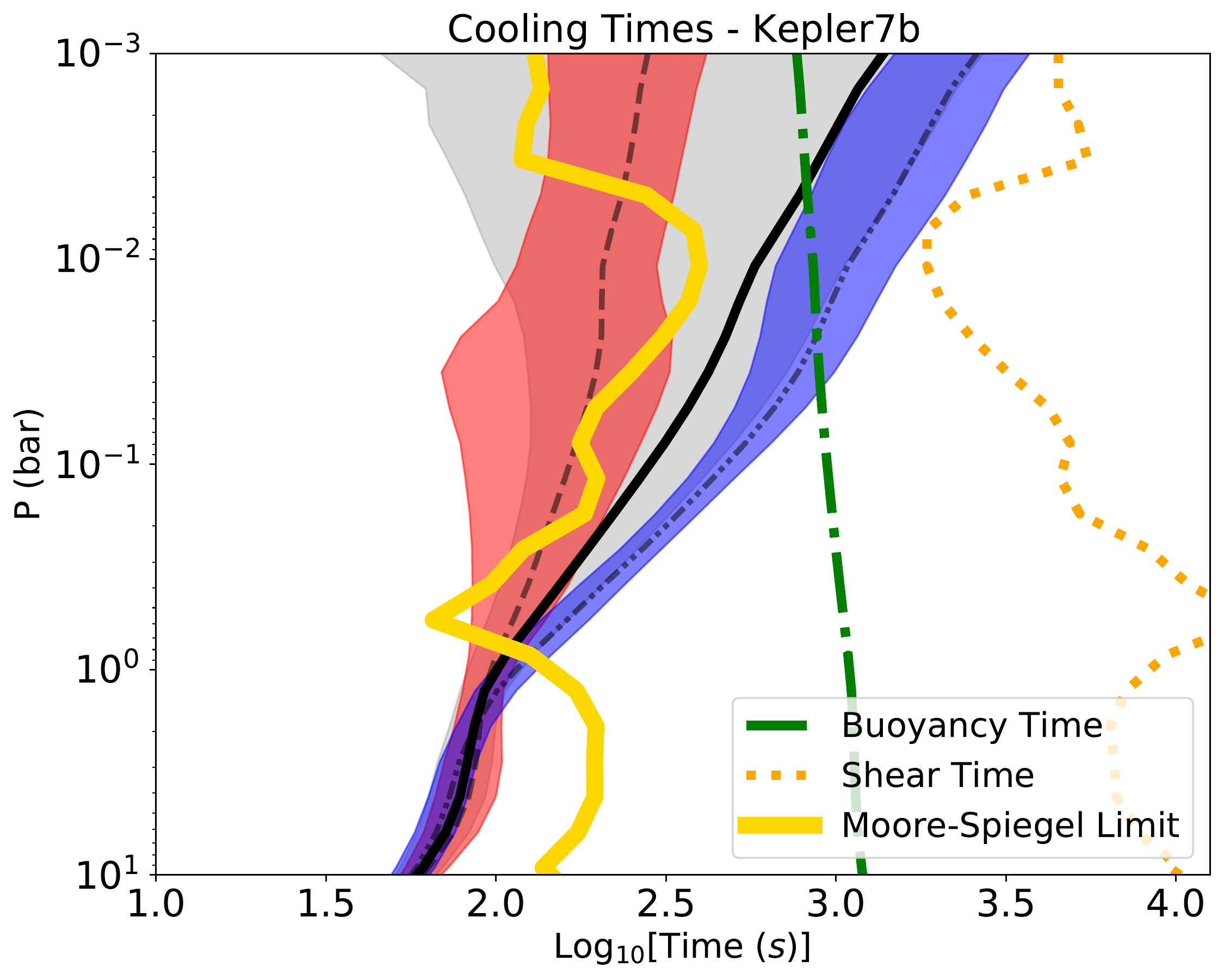}
    \caption{Vertical profiles of heat exchange ("cooling") times for transparent perturbations in the atmospheric models of HD189733b (top panel), HD209458b (middle) and Kepler7b (bottom). In each panel, three profiles are shown, each with an accompanying shaded region to highlight one standard deviation from the mean profile. The global average profile is shown as a solid line (with grey shading), while the day-side only and night-side only profiles are shown as dashed (with red shading) and dash-double-dot (with blue shading) lines,  respectively.  Cooling times are thus systematically shorter on the hot day-sides.  Vertical profiles of the buoyancy, shear and Moore-Spiegel times, all defined in the text, are also shown for comparisons(see legend for details). }
    \label{fig:times}
\end{figure}

\begin{figure}
	%\includegraphics[width=\columnwidth]{UZ_profile_noVD.pdf}
	\includegraphics[width=\columnwidth]{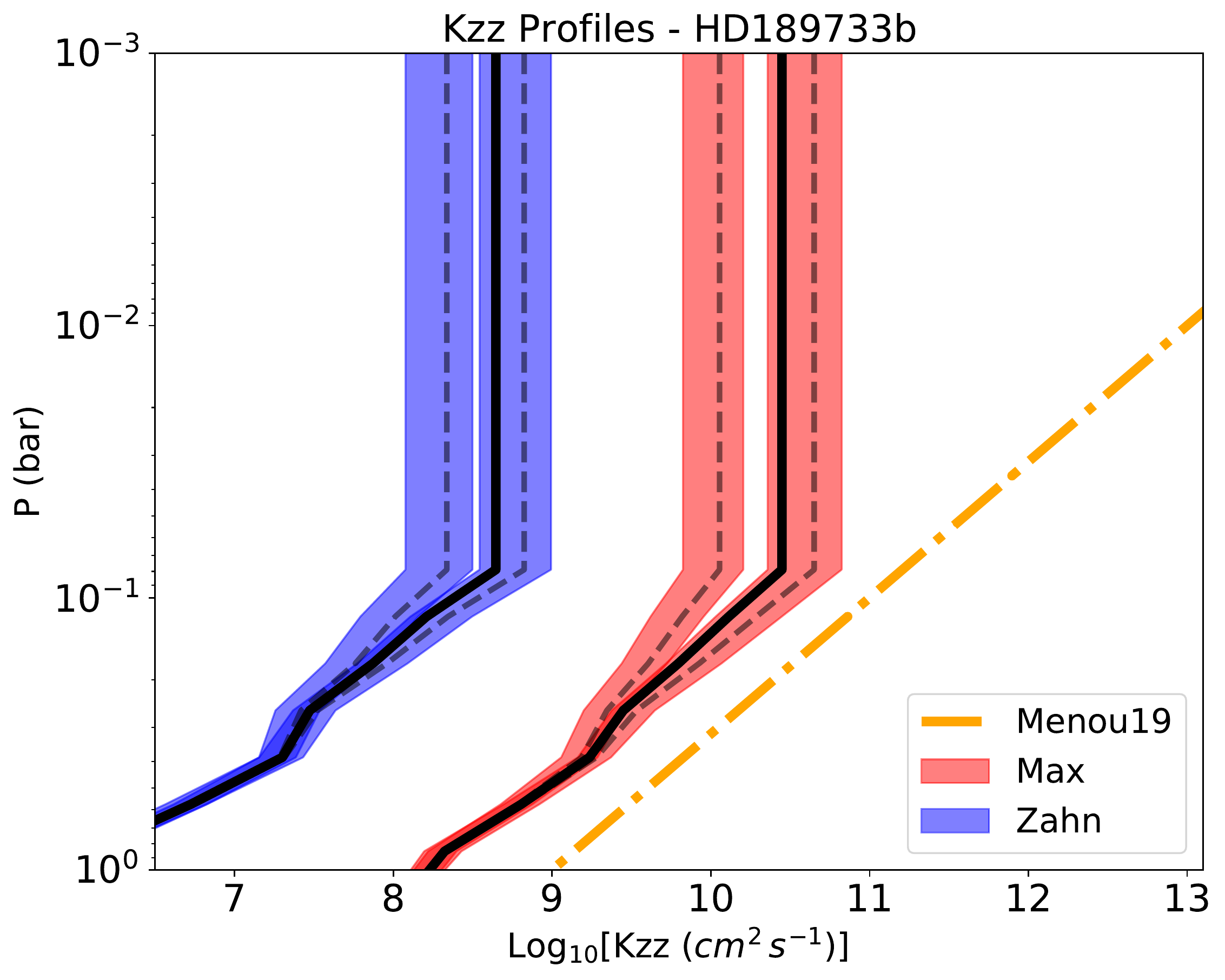}
	\includegraphics[width=\columnwidth]{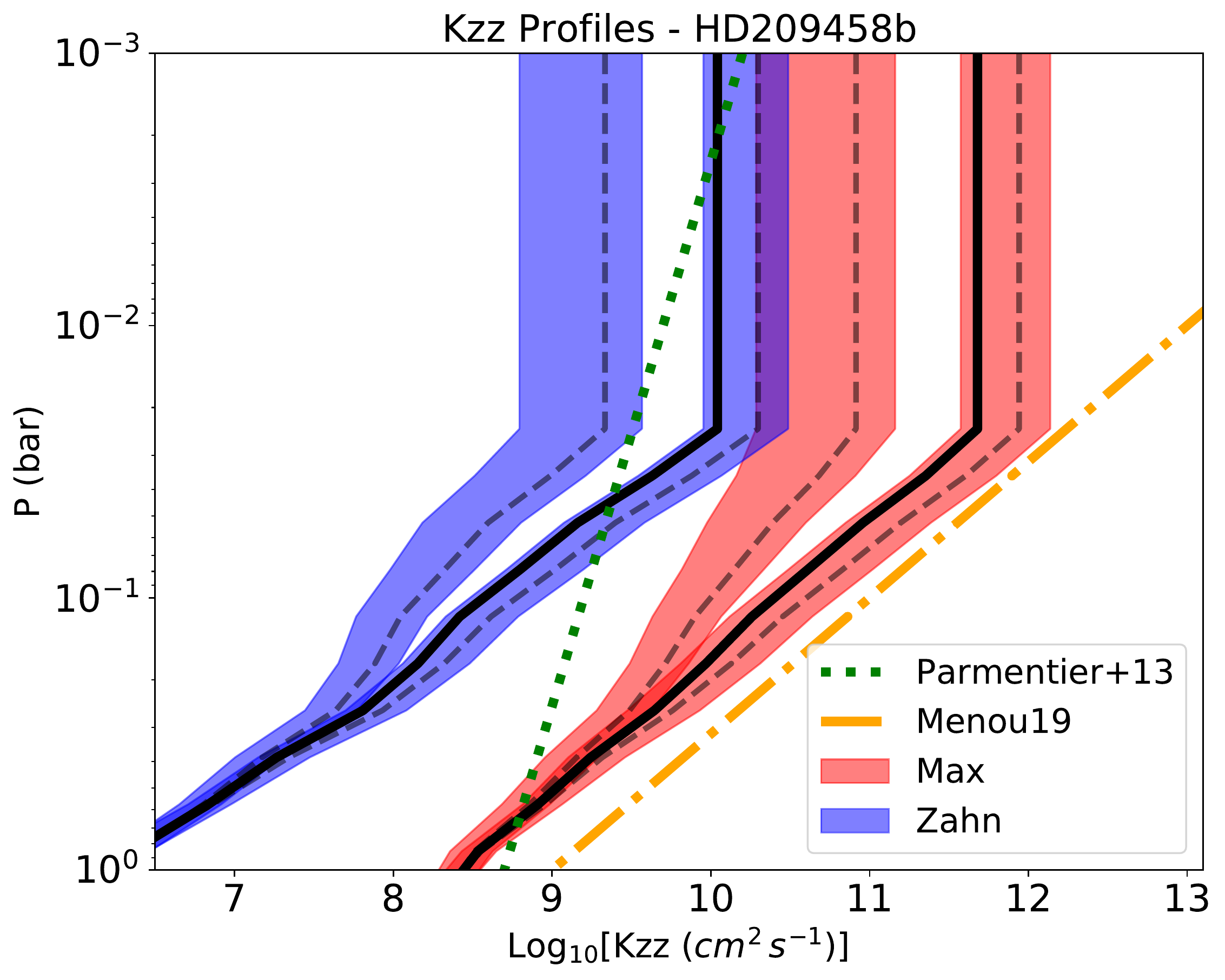}
	\includegraphics[width=\columnwidth]{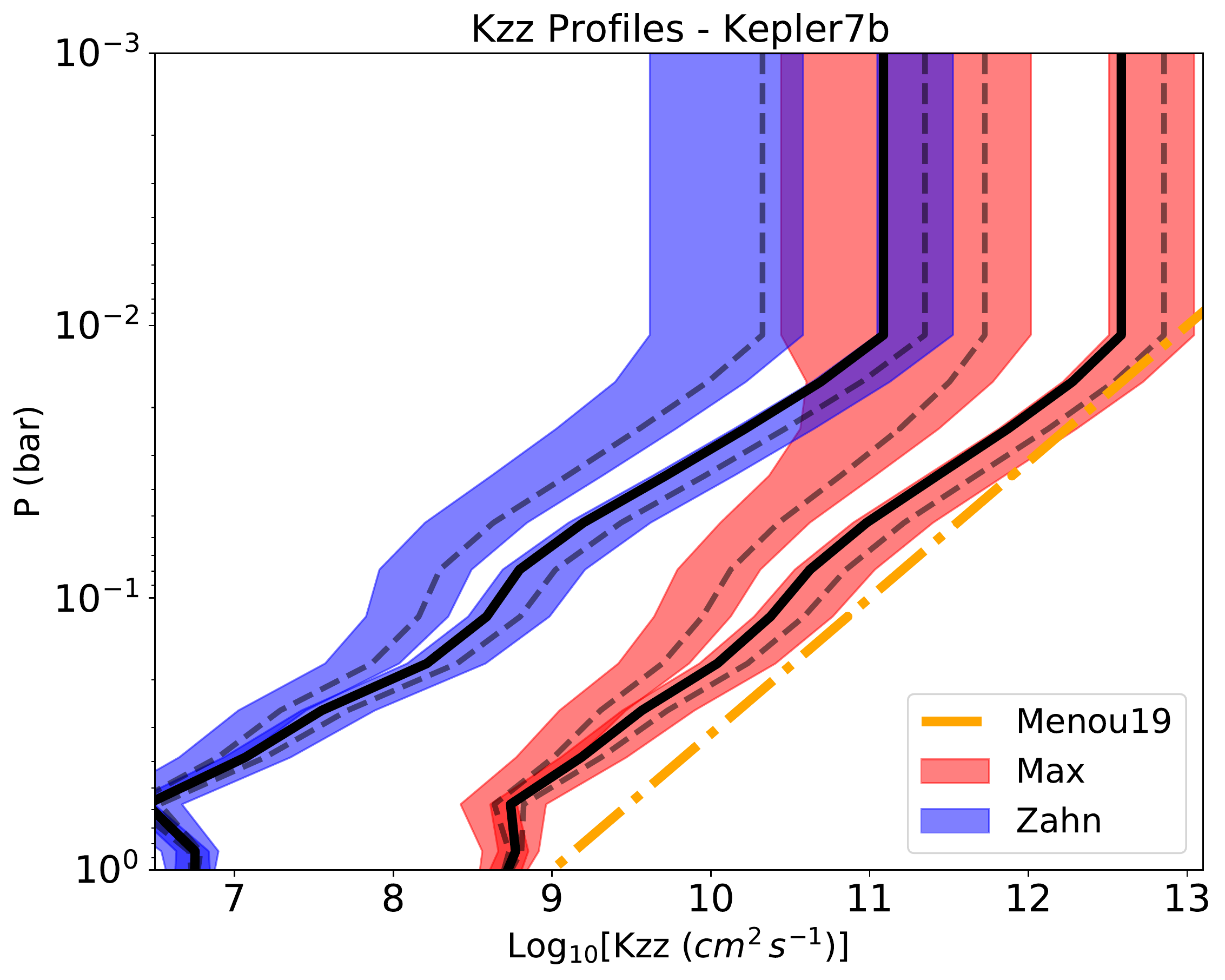}
    \caption{$K_{\rm zz}$ profiles for HD189733b (top panel), HD209458b (middle) and Kepler7b (bottom) computed in the double-diffusive (opaque) regime. The profiles for the Max transport formulation and Zahn transport formulation are shown with red and blue shaded regions, respectively. In each case, day-side averaged and night-side averaged profiles and a standard deviation (shaded area) are shown. The black solid line is a profile averaged over the entire atmosphere. For comparison, the profile discussed by \protect\cite{2013A&A...558A..91P} (green dotted line in the top panel) and the profile used by \protect\cite{2019MNRAS.485L..98M} (orange dash-dotted lines) are also shown. }
    \label{fig:Kzz}
\end{figure}

\begin{figure}
	%\includegraphics[width=\columnwidth]{UZ_profile_noVD.pdf}
	\includegraphics[width=\columnwidth]{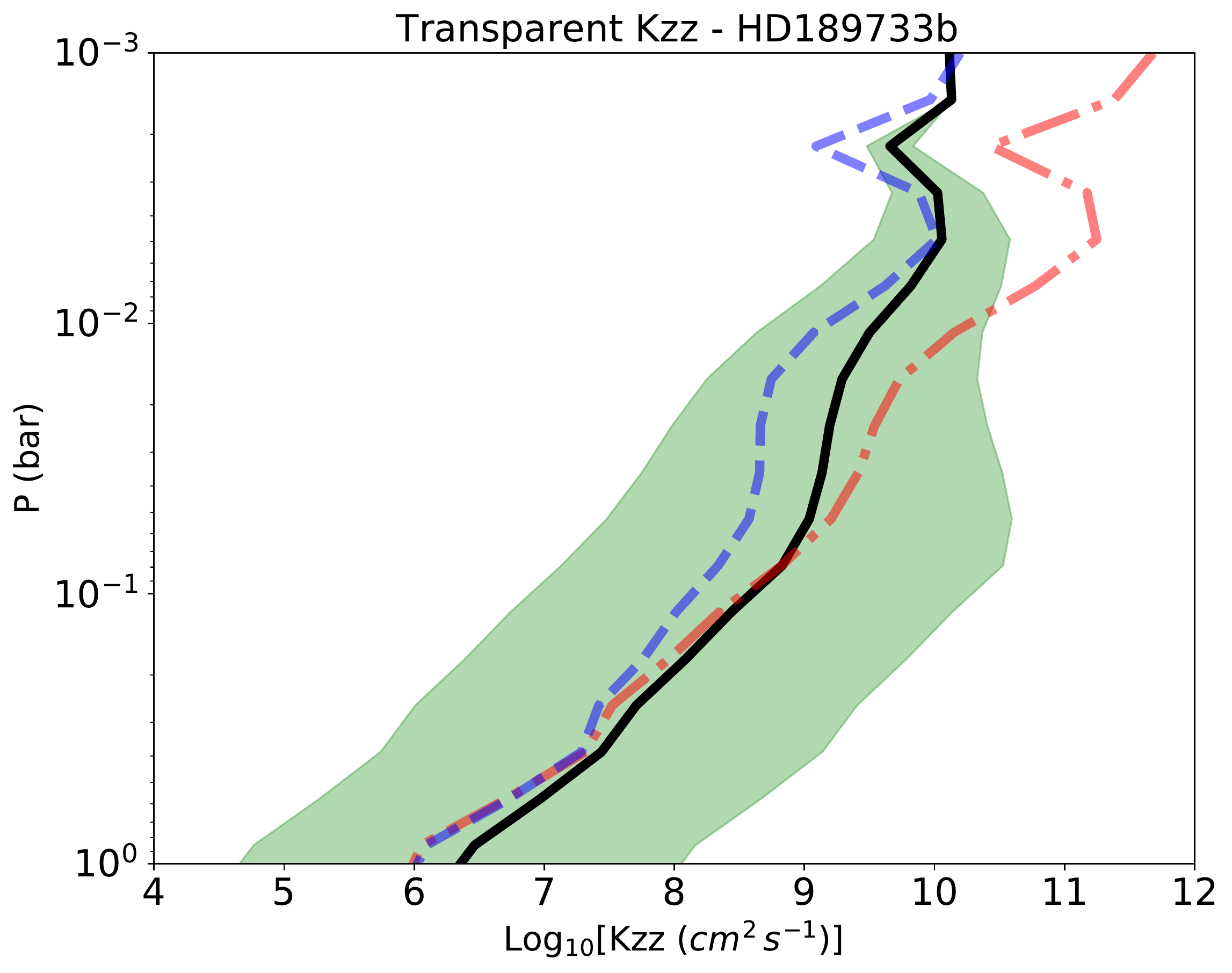}
	\includegraphics[width=\columnwidth]{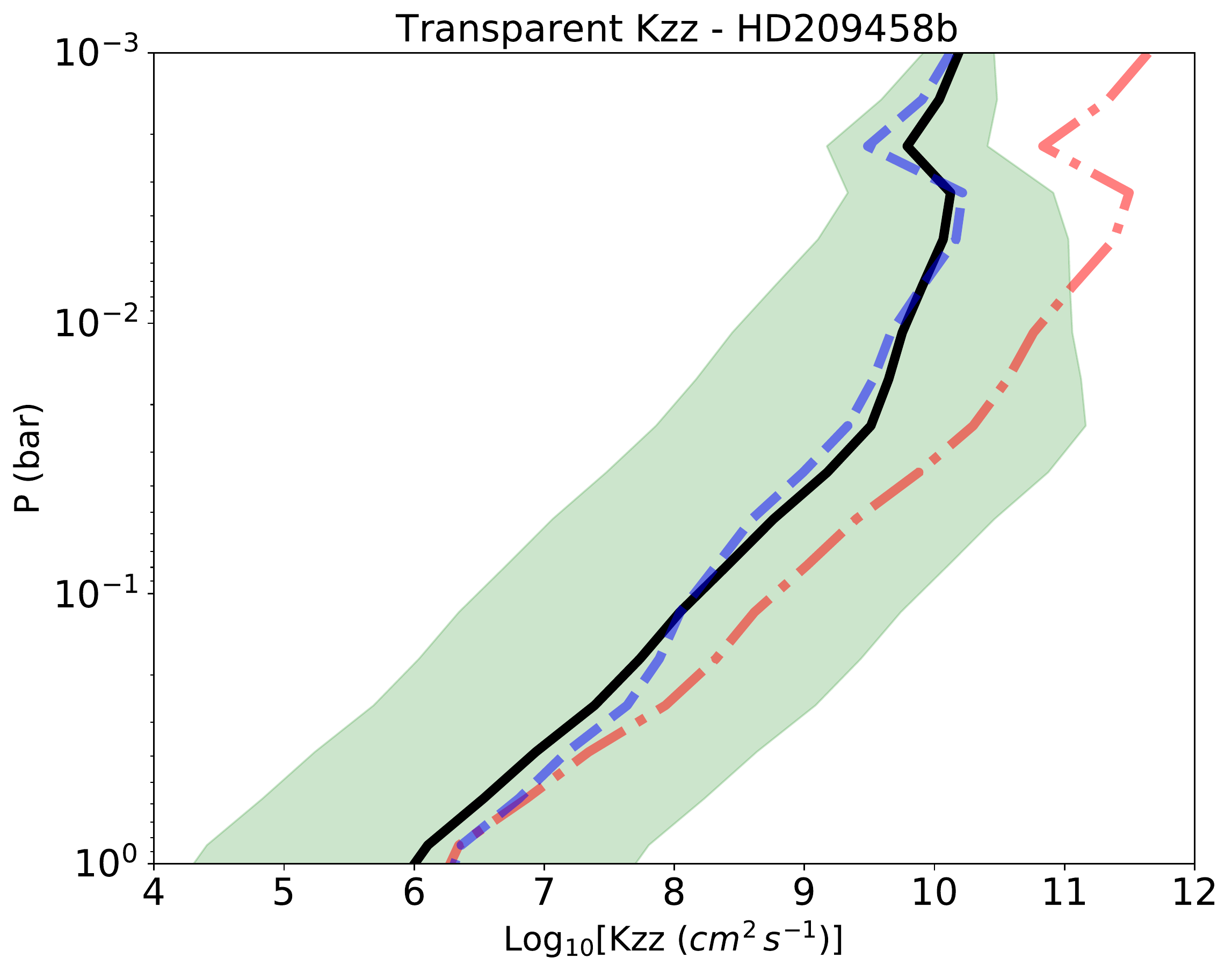}
	\includegraphics[width=\columnwidth]{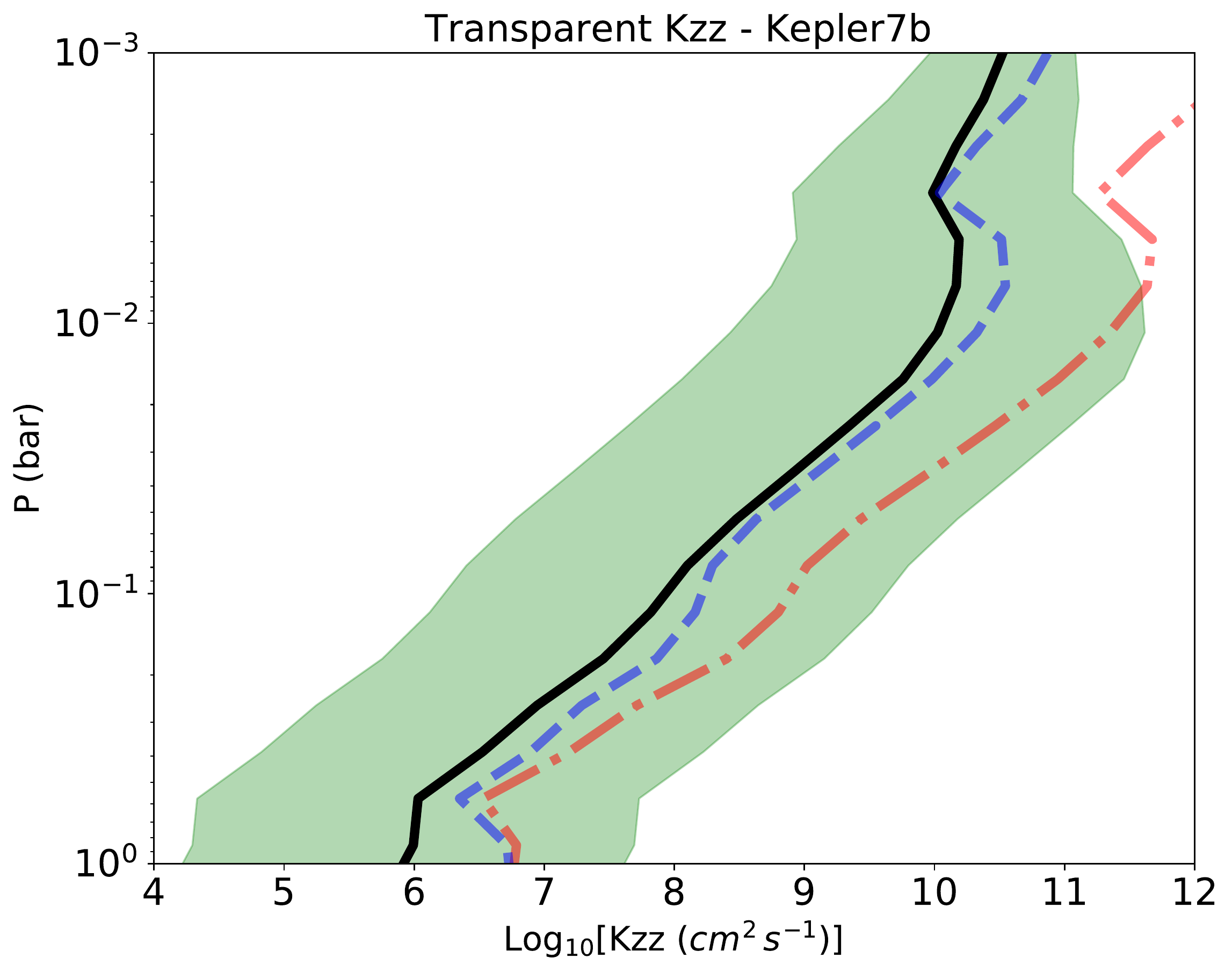}
    \caption{Profiles of $K_{\rm zz}^{\rm thin}$ computed in the semi-transparent regime, for our three planets of interest. The solid line shows a logarithmic-average of the two bounding profiles that define the green shaded region. For comparison, the blue dashed and red dash-dotted lines reproduce the night-side and day-side $K_{\rm zz}^{\rm Zahn}$ profiles shown in Fig.~8, respectively.  The shaded region gets narrower at low-pressures as a limiting turbulence integral scale of $0.3 \, H_p$ is imposed.   }
    \label{fig:Kzz_transparent}
\end{figure}

Figure~\ref{fig:Kzz} shows vertical $K_{\rm zz}$ profiles for our three planets of interest, comparing $K_{\rm zz}^{\rm Zahn}$ and $K_{\rm zz}^{\rm Max}$ prescriptions. For each $K_{\rm zz}$ prescription, global, day-only and night-only profiles are shown with their respective standard deviations. In addition, at low enough pressures, where the scale $\lambda_{\rm buoy}$ approaches $H_{\rm p}$, the   $K_{\rm zz}$ profiles have been frozen to the value they take at the point where  $\lambda_{\rm Ross} = H_{\rm p}$ (at the boundary between the semi-transparent and opaque regimes). For comparison, the analytic $K_{\rm zz}$ profile used in \cite{2019MNRAS.485L..98M}, 
\begin{equation}
K_{\rm zz}  =  10^9 \left( \frac{P}{1~{\rm bar}} \right)^{-2}~{\rm cm^2 ~s^{-1}},
\end{equation} 
is also shown as a dash-dotted (orange) line, and the $K_{\rm zz}$ profile inferred by \cite{2013A&A...558A..91P} is shown as a dashed (green) line for HD209458b only.

These $K_{\rm zz}$ profiles are only indicative because they are subject to substantial uncertainties. Below photospheric levels, the slanted portions of these profiles assume the optically-thick limit, even though the turbulent medium is semi-transparent on turbulent eddy scales well below the integral scale (see Fig.~\ref{fig:length})  The location of the transition to a fully optically-thin regime of secular shear turbulence (i.e.  the kink in the $K_{\rm zz}$ profiles of Fig~\ref{fig:Kzz}) is uncertain itself because the integral scale and the transition to transparent fluctuations are not precisely known (see shaded areas in Fig.~\ref{fig:length}). Finally, the transparent secular shear turbulence regime thought to be relevant at and above photospheric levels (vertical portions above the kink in Fig.~\ref{fig:Kzz}) makes it likely that the minimal heat exchange time arguments discussed in connection to Fig.~\ref{fig:times} above become relevant. Optically-thin regions with long enough heat exchange times could remain stable to secular shear turbulence, which would result in no enhanced transport ($K_{\rm zz} \to 0$).  This could be the case on the nightsides of all three planets considered, and even on the dayside of HD189733b.   

For atmospheric regions where transparent shear turbulence is expected (e.g. the dayside of Kepler7b, at and above photospheric levels),  we would like to have a scaling theory for the magnitude of transport that goes beyond a crude vertical extrapolation of the double-diffusive (optically-thick) regime.  We now turn to such an improvement by deriving a scaling that accounts for the transparency of the medium.

\subsubsection{Transparent Regime: A Heuristic Scaling}

To the best of our knowledge, no turbulence model exist for the transparent and semi-transparent regimes of secular shear turbulence relevant to photospheric levels in hot Jupiter atmospheres.  Nevertheless, it is possible to consider a heuristic scaling based on a few determining physical principles.

In a semi-transparent turbulent regime, with optically-thick large-scale fluctuations and optically-thin small-scale fluctuations,  we expect the opacity transition scale, from opaque to transparent, to play a special role.  It corresponds to the largest fluctuation scale still subject to the shortest heat exchange time available in the turbulent medium.   That is because heat exchange occurs on a shorter diffusion time as one goes from large opaque scales to smaller (but still opaque) scales, until the scale-free regime of a constant heat exchange time is reached as the medium becomes transparent, at the opacity transition scale.  On transparent (small enough) scales,  there is no particular advantage in going to smaller scales to achieve faster heat exchange.

We can combine this property with a general expectation that a shear instability can also grow in scale-free manner (modulo the action of viscosity, which limits its growth on the smallest scales).  Assuming a growth rate equal to the shear timescale and a largest unstable scale set by the opacity transition scale, this suggests a magnitude of turbulent transport set by by 'largest mode' available in the transparent regime, or
\begin{equation}
K_{\rm zz}^{\rm thin} \sim S \Lambda^2,
\end{equation}
where $\Lambda$ might be chosen as a value between the bounds $\lambda_{\rm ph}$ and $\lambda_{\rm Ross}$ shown in Fig.~6 (depending on our specific choice of opacity transition scale).   

In this simple formulation,  smaller transparent scales may be (nearly) equally unstable but would contribute minimally to the effective turbulent transport (which is dominated by the largest transparent scale $\Lambda$), while larger opaque scales would experience longer heat exchange times (in the double-diffusive regime) and thus contribute minimally, or not at all, in driving the secular shear turbulence.
   
Figure~\ref{fig:Kzz_transparent} shows profiles of $K_{\rm zz}^{\rm thin}$ computed with Eq.~18, for our three planets of interest. In each panel, the shaded region is bounded by using either the smaller $\lambda_{\rm ph}$  or the larger $\lambda_{\rm Ross}$ in our definition of $K_{\rm zz}^{\rm thin}$. The shear rate $S$ is obtained from the diagnostic shear profile (globally-averaged) computed from our atmospheric models. The solid line shows a logarithmic-average profile of the two bounding profiles (defining the shaded region). For comparison, the dashed and dashed-dotted lines show, respectively,  the night-side and day-side $K_{\rm zz}^{\rm Zahn}$ profiles in each atmosphere.  Note that the shaded region gets narrower at low-pressure as we limit the maximum value of the length-scale $\Lambda$ to the maximum integral scale of $0.3 \, H_p$, as discussed in \S3.1.1.     
 
Figure~\ref{fig:Kzz_transparent} shows results for globally-averaged profiles of density, which directly enters the definition of $\lambda_{\rm ph}$ or $\lambda_{\rm Ross}$.  At a given pressure level, we find atmospheric densities that can vary by a factor of several from the day-side to the night-side, so that corresponding variations of $K_{\rm zz}^{\rm thin}$ of an order of magnitude, or more,  can generally be expected (with stronger transport on the lower density dayside). Opacity effects could further impact these differences in magnitude of transport between the day- and night-sides.

In summary,  the solid line profiles in  Figure~\ref{fig:Kzz_transparent} constitute our best estimate (with errorbars indicated by the green shaded region) of the magnitude of turbulent transport to be expected from secular shear instabilities in the (semi-)transparent regime.  While these profiles coincidentally lie close to the $K_{\rm zz}^{\rm Zahn}$  profiles derived from the optically-thick theory, we recommend using the semi-transparent scaling as it is physically better motivated.

Our various profiles should always be considered in concert with the heat exchange times requirements shown in Fig. ~\ref{fig:times}, since secular instabilities and the resulting turbulent transport are not expected to be present unless short enough heat exchange times are realized in the atmosphere of interest.  We chose to show all $K_{\rm zz}$ profiles in Fig.~8 and~9 irrespective of any specific heat exchange time criterion because the threshold for fast enough heat exchange is itself rather uncertain (as discussed in \S3.1.2).

\subsection{Turbulent Feedback on the Mean Atmospheric Flow
\label{sec:feedback} }

Another potential source of uncertainty for secular shear turbulence in hot Jupiter atmospheres is the possibility that the resulting turbulent transport of momentum ends up modifying the mean atmospheric flow \citep{2019MNRAS.485L..98M}. This is a typical outcome of shear instabilities, which act to neutralize the unstable shear profiles that generated them in the first place. We have found that this source of uncertainty, which potentially requires dynamically coupling shear turbulent transport to feedback on the mean flow computation, is fairly minor for the three planets of interest here.

To quantify the degree of feedback,  we have performed the following numerical exercise: we ran two models in succession, first without and then with the coupling between vertical transport and the mean flow included.  In the first model,  we ran our model to completion without any vertical transport applied on the flow.  This transport-free flow is used diagnostically to compute the magnitude of the turbulent vertical transport expected.  The globally-averaged profiles of $K_{\rm zz}$ that result are then fed into a second set of models in which the turbulent transport is applied in both the heat and momentum equations for the atmospheric flow (see Appendix A for details).  By the end of the second set of models,  we are in a position to compute a new set of 'iterated' $K_{\rm zz}$ profiles (with feedback accounted for) and can compare them to the profiles obtained from our first set of no-feedback models.  In all cases,  we find that the 'iterated' $K_{\rm zz}$ profiles are similar to the no-feedback $K_{\rm zz}$ profiles. This leads us to conclude that momentum feedback on the atmospheric flow is minor to negligible.  Appendix~C provides the quantitative results of this analysis.

We note that this suggests, as a corollary,  that it is a reasonable approach to add $K_{\rm zz}$ contributions that arise from the global circulation pattern \citep{2013A&A...558A..91P,2019ApJ...881..152K}  to those from the turbulent vertical transport discussed in the present work. In other words, the strongest contribution from these two distinct processes would simply dominate in regions where they co-exist \citep[see also][]{2018ApJ...866....1Z,2018ApJ...866....2Z}

\section{Conclusions}

We have discussed the nature and magnitude of vertical transport from semi-transparent secular shear instabilities in hot Jupiter atmospheres. Our analysis relies on numerical simulations of the atmospheric flows on HD209458b,  HD189733b and Kepler7b,  covering a range of surface gravities and radiative equilibrium temperatures. 

A key take away from our analysis is that the presence and magnitude of turbulent vertical transport ($K_{\rm zz}$ ) in the atmospheres of these three hot Jupiters are all uncertain.  At face value,  our work suggest that no significant vertical transport from shear turbulence is expected on a cool planet like HD189733b.  Day side vertical transport may be expected on warmer hot Jupiters such as HD209458b,  while both day-side and night-side vertical transport may occur on hot Jupiters as hot or hotter than Kepler7b.  In all cases,  vertical transport is somewhat favoured in equatorial regions and disfavoured at mid-latitudes and in polar regions.  Such heterogeneous mixing, especially the strong day vs night regime differences,  is not easily captured with 1D atmospheric models.

Given the large uncertainties inherent to our analysis,  a more lenient interpretation would allow for turbulent vertical transport everywhere on the three planets studied,  except on the cool  nightside of HD189733b.  By contrast, a more conservative interpretation would restrict any turbulent vertical transport to the day side of Kepler7b.  In all cases, our best estimate of the magnitude of secular shear turbulent transport (when present) is derived from a semi-transparent scaling (\S3.2.2),  which is subject to large uncertainties itself, as shown in Figure~\ref{fig:Kzz_transparent}.

The almost qualitative level of uncertainty affecting our conclusions is clearly not desirable.  On the bright side,  the situation offers many opportunities for progress via comparative studies across the spectrum of hot Jupiters with different equilibrium temperatures and by comparing the day and night sides on any given planet. Another avenue for progress may become possible by comparison with isolated brown dwarfs and young luminous Jupiters.  Lacking the strong instellation that characterizes hot Jupiters,  one may expect considerably slower atmospheric winds and correspondingly reduced shear rates on these self-luminous objects. This might significantly reduce the role of shear-driven turbulent vertical transport on these objects,  even at atmospheric temperatures and surface gravities such that it does affect hot Jupiter atmospheres.

%%%%%%%%%%%%%%%%%%%%%%%%%%%%%%%%%%%%%%%%%%%%%%%%%%

%%%%%%%%%%%%%%%%%%%%%%%%%%%%%%%%%%%%%%%%%%%%%%%%%%

\section*{Data Availability Statement}
The data underlying this article will be shared on reasonable request to the corresponding author.

\section*{Acknowledgements}
The author thanks Jeremy Goodman for advice on the existing literature regarding secular shear stability in the optically-thin limit and an anonymous referee for comments that helped improve the quality of the manuscript.
Computing resources were provided by the Canadian Institute for Theoretical Astrophysics at the University of Toronto. KM is supported by the National Science and Engineering Research
Council of Canada. This work has made extensive use of the following software packages: {\tt matplotlib}.

We would furthermore like to acknowledge that this work was performed on land which for thousands of years has been the traditional land of the Huron-Wendat, the Seneca and the Mississaugas of the Credit. Today this meeting place is still the home to many indigenous people from across Turtle Island and we are grateful to have the opportunity to work on this land.

%%%%%%%%%%%%%%%%%%%%%%%%%%%%%%%%%%%%%%%%%%%%%%%%%%



\bibliographystyle{mnras}
\bibliography{uncertain_kzz}

%%%%%%%%%%%%%%%%% APPENDICES %%%%%%%%%%%%%%%%%%%%%

\appendix

\

\section{PlaSim-Gen Vertical Transport Scheme}

As detailed in PlaSim's user manual \citep{fraedrich2005}, vertical transport of momentum and heat is implemented by solving implicitly the following vertical diffusion equations in each atmospheric column of the model
\begin{eqnarray}
\frac{\partial u}{\partial t} &=& \frac{1}{\rho} \frac{\partial }{\partial z} \left(  \rho K_{\rm m} \frac{•\partial u}{\partial z} \right)\\
\frac{\partial T}{\partial t} &=& \frac{1}{\rho} \frac{\partial }{\partial z} \left(  \rho K_{\rm h}   (\frac{p}{p_s})^{R_d/C_p}  \frac{•\partial \theta}{\partial z} \right),
\end{eqnarray}
where $\rho$ is the mass density, $p_s$ is the local surface pressure and $\theta$ is the potential temperature (a proxy for entropy).  In PlaSim's original version,  designed to model Earth's climate,  the transport coefficients $K_{\rm m}$ and $K_{\rm h}$ are computed from a mixing length approach focused on surface exchange coefficients relevant to the planetary boundary layer (i.e. describing ground-atmosphere exchanges).  We have modified the original formulation to replace the dynamically calculated values of $K_{\rm m}$ and $K_{\rm h}$  with steady vertical profiles of transport coefficient values (taken as model parameters). The original top and bottom zero-flux boundary conditions were kept untouched.

\section{Upper Boundary Effects}

We carried out a vertically extended simulation of the atmospheric flow on HD209458b to evaluate the effect of the model upper boundary location on the simulated flow,  in particular the increased shear rate found near all model tops (as shown in Figure 5).  To maintain comparable
vertical resolution in the original and in the vertically extended model,  we extend the atmosphere to a pressure level $P_{\rm top} = 10^{-5}$~bars and run the extended model at greater T31L42 resolution (vs. T31L31 originally).  The model parameter OOM$=7$ was adopted to reach $10^{-5}$~bars at the top (keeping the same bottom pressure).  The time step was reduced
by a factor of 2 (MPSTEP $= 90$s,  from $180$s) to guarantee numerical convergence.

Figure B1 shows the zonally-averaged zonal wind profile for the vertically extended model of HD209458b.  The peak wind location and magnitude are comparable in both models ($\sim 10^{-1}$~bar,  $\sim 4300$~m/s),  as does the general zonal wind pattern.  Stronger retrogrades winds on the flank of the equatorial jet are noticeable near the model top boundary in the vertically extended model.

Figure B2 shows vertical profiles of thermal buoyancy and shear rate in the vertically extended model,  to be directly compared to the top panel of Figure~4.  The excess shear at the $10^{-2}$~bar level in the original model  is now found at the  $10^{-4}$~bar level in the vertically extended model.  This is strongly suggestive that the upper model results are affected by the proximity of the model top boundary.  By contrast, the bulk of the atmospheric domain displays a well-behaved shear profile,  which is the focus of our analysis.

For completeness,  Figure B3 shows vertical profiles of heat exchange and other relevant times discussed in our analysis.  This figure is to be directly compared to the top panel of Figure 7.  The general consistency between the various profiles in each case (vertically-extended vs original) indicates that the main results of our analysis are not strongly affected by upper boundary effects affecting the shear rate.

\begin{figure}
	%\includegraphics[width=\columnwidth]{UZ_profile_noVD.pdf}
	\includegraphics[width=\columnwidth]{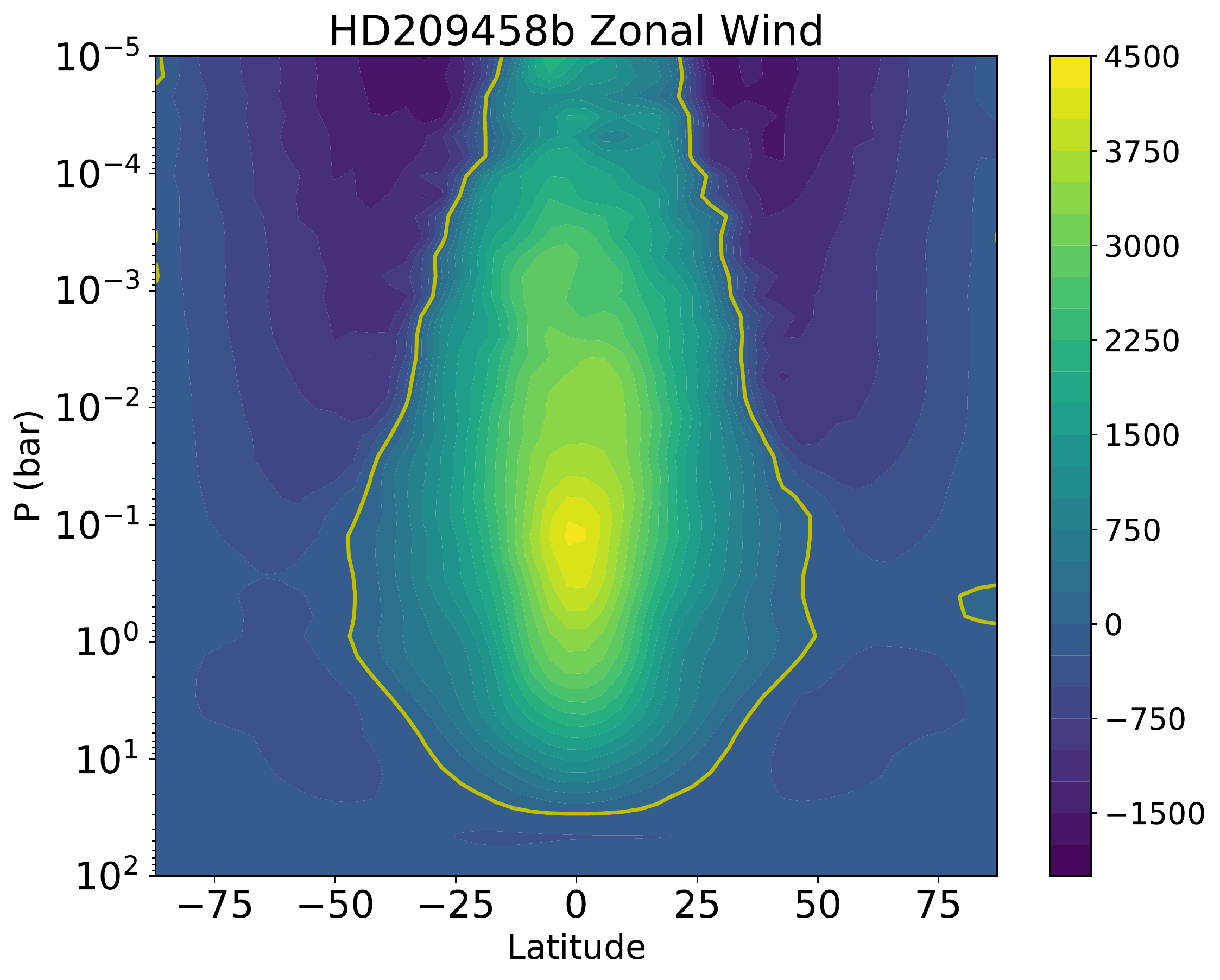}	
    \caption{Vertically extended version of the zonally-averaged zonal wind profile at planet day 600 in our T31L42 model of HD209458b (with P$_{\rm top}=10^{-5}$~bar). Zonal wind speeds are slightly higher than in our T31L30 model. The counter-jets at mid-latitudes are stronger and peak near the (extended) model top. Overall, the zonal wind and its vertical shear are consistent with the vertically less extended T31L30 model.}
    \label{fig:T42_UZ}
\end{figure}

\begin{figure}
	%\includegraphics[width=\columnwidth]{UZ_profile_noVD.pdf}
	\includegraphics[width=\columnwidth]{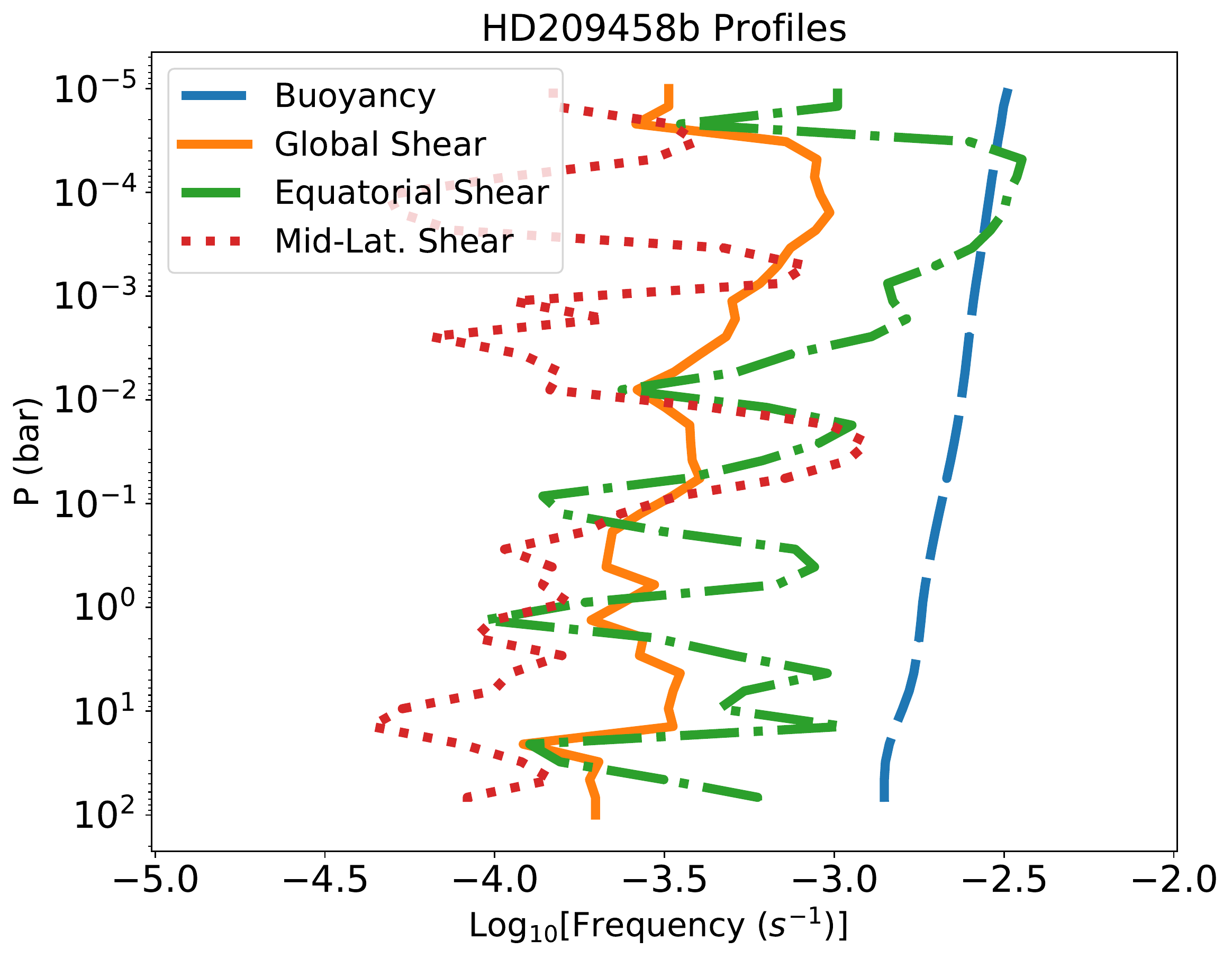}	
    \caption{Vertically extended version of vertical profiles of buoyancy frequency, $N$,  and shear frequency, $S$,  in our T31L42 model of HD209458b (with P$_{\rm top}=10^{-5}$~bar). Excess shear near the model top has moved up at about the $\sim 10^{-4}$~bar level.  This excess shear is likely caused by boundary effects.}
    \label{fig:T42_shear}
\end{figure}

\begin{figure}
	%\includegraphics[width=\columnwidth]{UZ_profile_noVD.pdf}
	\includegraphics[width=\columnwidth]{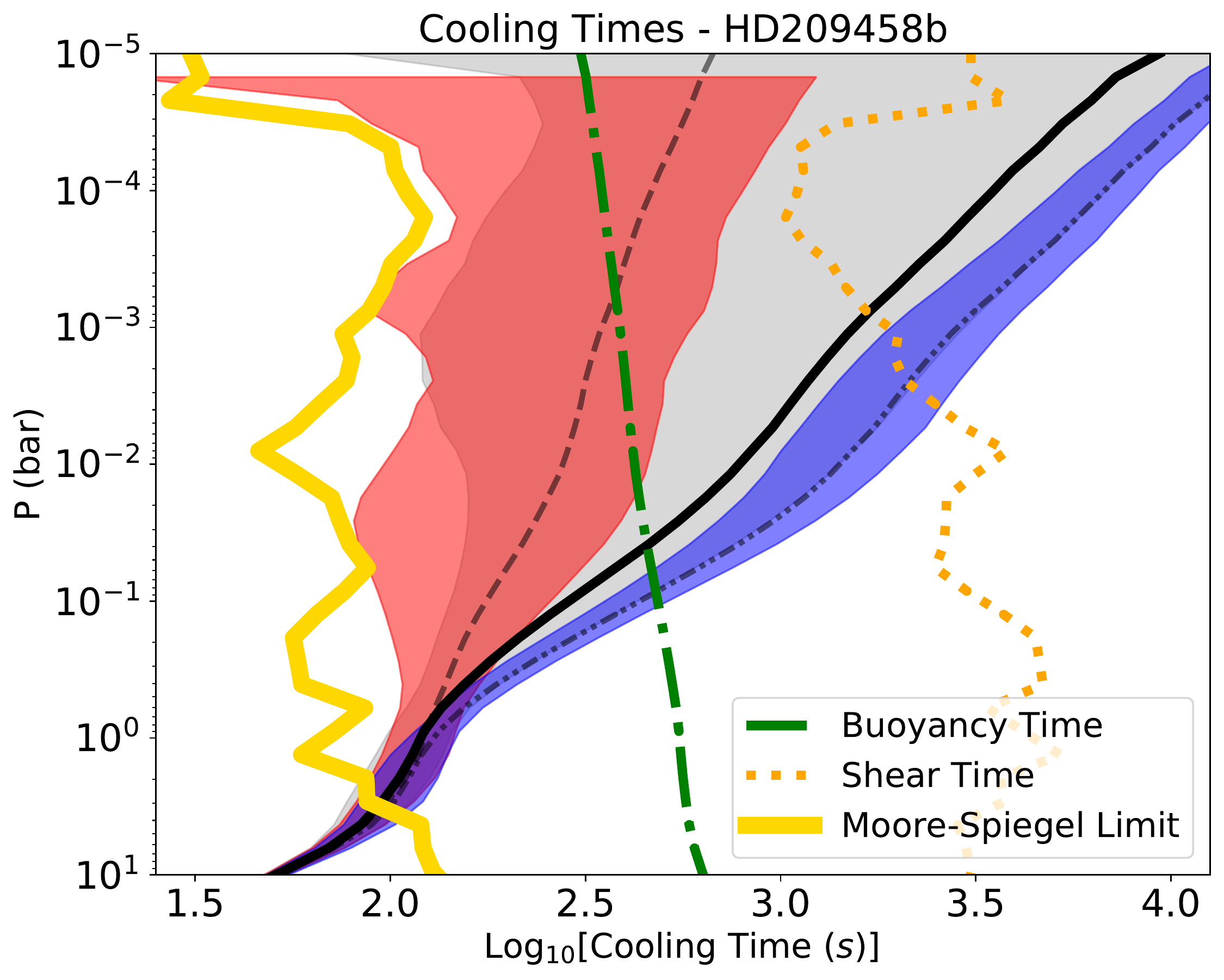}	
    \caption{Vertically extended version of vertical profiles of cooling times  in T31L42 model of HD209458b (with P$_{\rm top}=10^{-5}$~bar). The various profiles shown are consistent with the original T31L30 versions shown in the top panel of Figure~7.}
    \label{fig:T42_times}
\end{figure}

\section{No-feedback vs.  Iterated Kzz Profiles}

\begin{figure}
	%\includegraphics[width=\columnwidth]{UZ_profile_noVD.pdf}
	\includegraphics[width=\columnwidth]{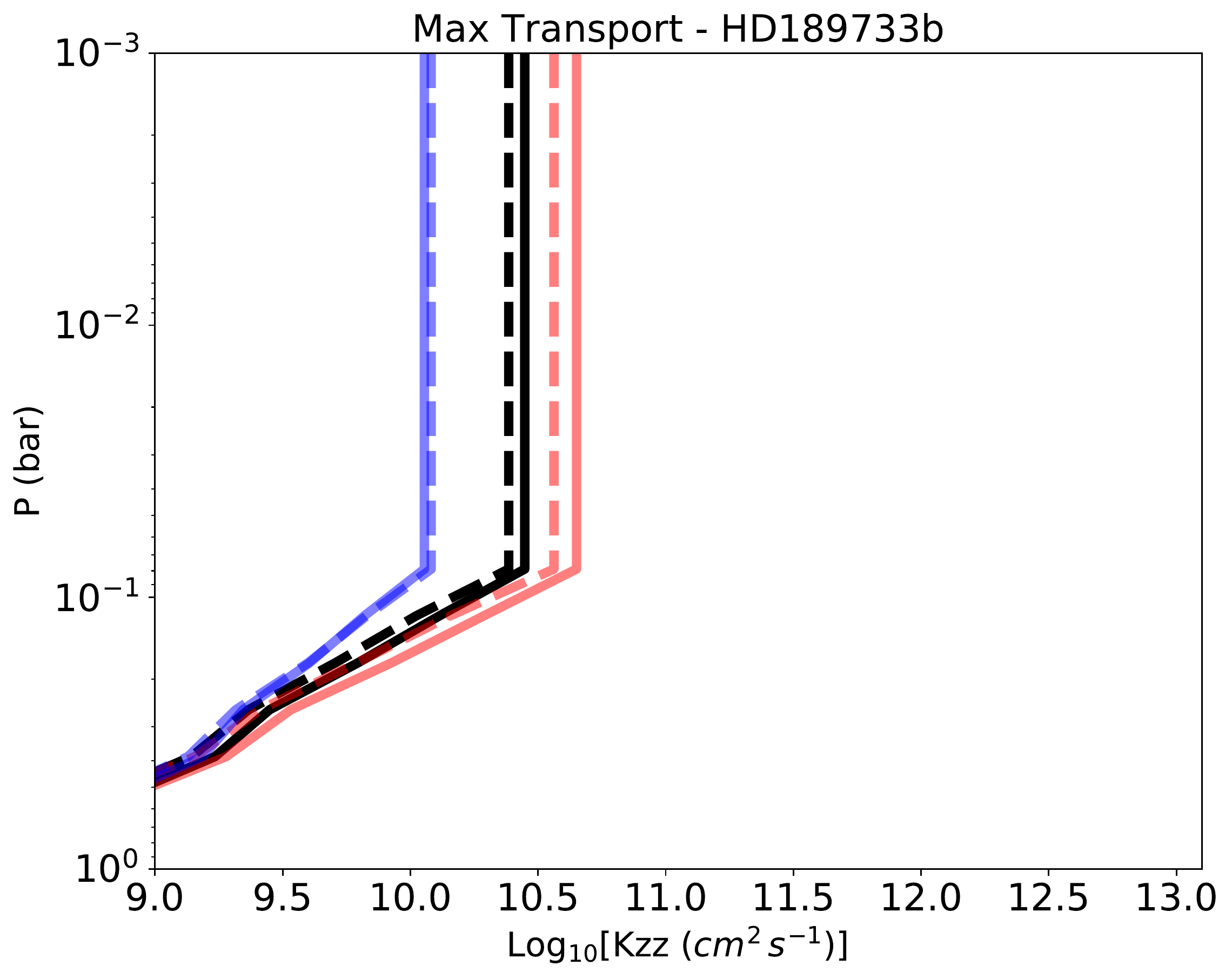}
	\includegraphics[width=\columnwidth]{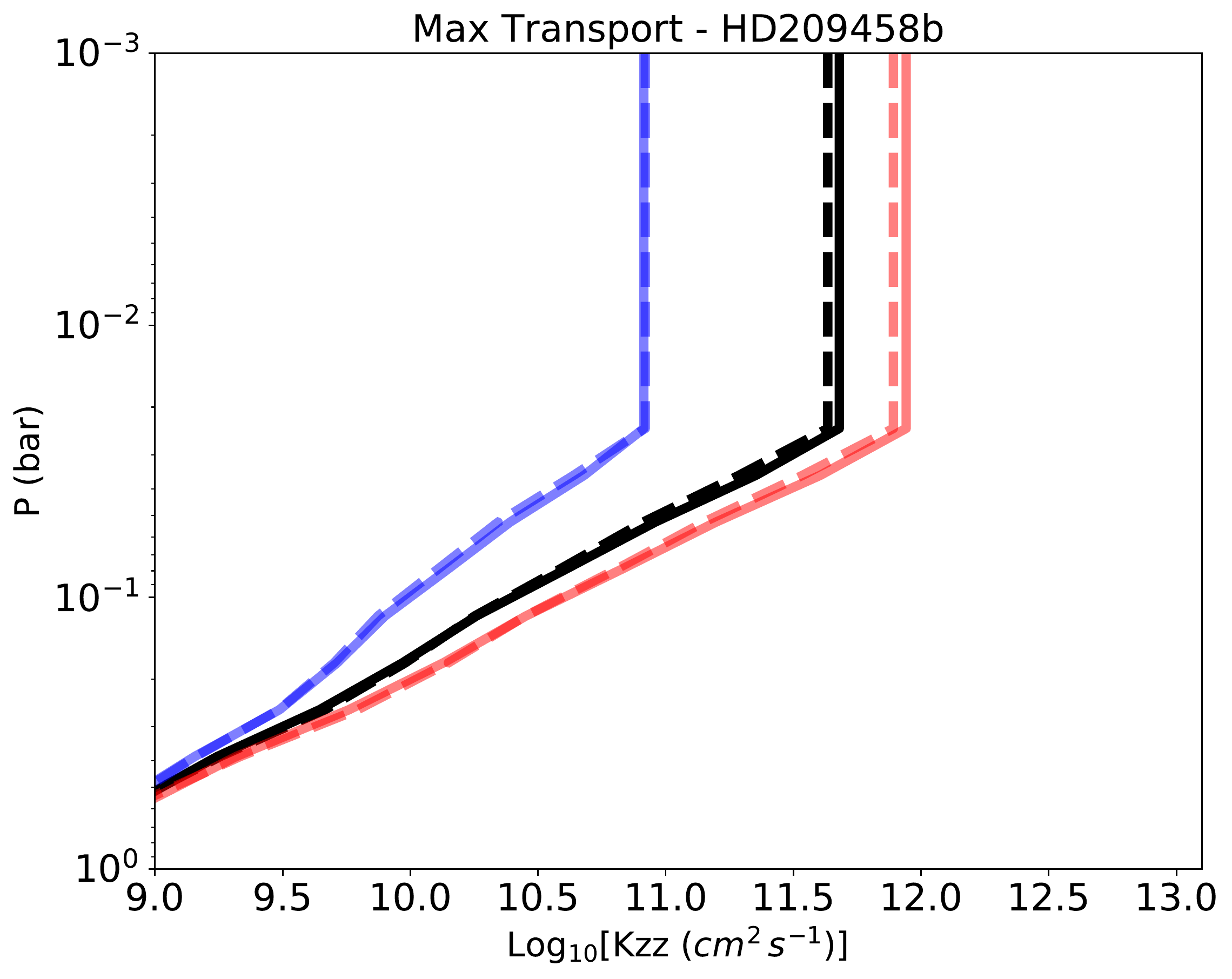}
	\includegraphics[width=\columnwidth]{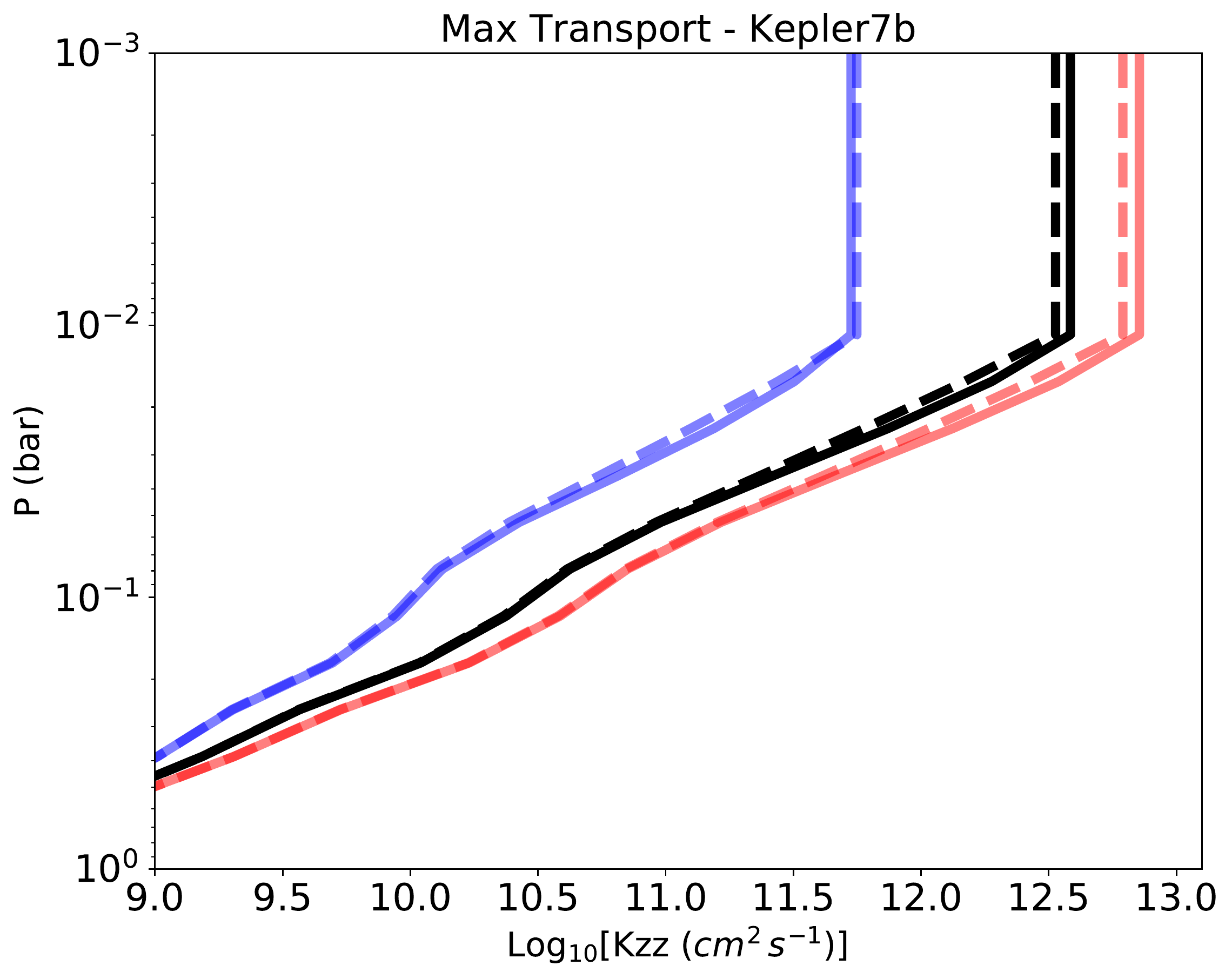}
    \caption{$K_{\rm zz}$ profiles according to the Max transport formulation for HD189733b (top panel), HD209458b (middle) and Kepler7b (bottom).  In each panel, day-averaged (red), night-averaged (blue) and globally-averaged (black) profiles are shown. Dashed profiles are non-iterated while solid profiles are iterated, as described in \S~3.3. The minor differences between profiles for iterated vs non-iterated flows suggest that turbulent transport feedback on the atmospheric flow can be ignored in first approximation.}
    \label{fig:Kzz_Max_iterated}
\end{figure}

\begin{figure}
	%\includegraphics[width=\columnwidth]{UZ_profile_noVD.pdf}
	\includegraphics[width=\columnwidth]{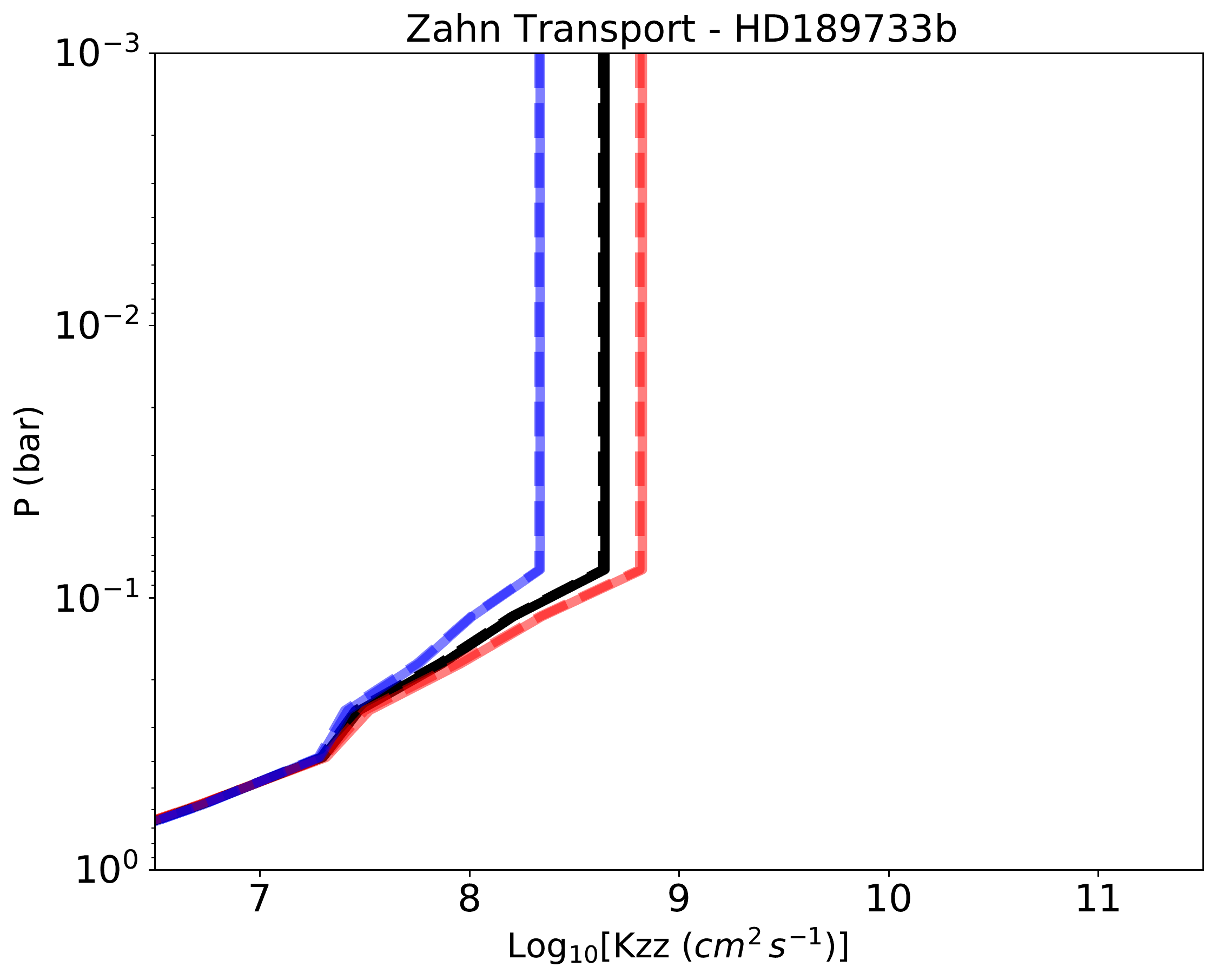}
	\includegraphics[width=\columnwidth]{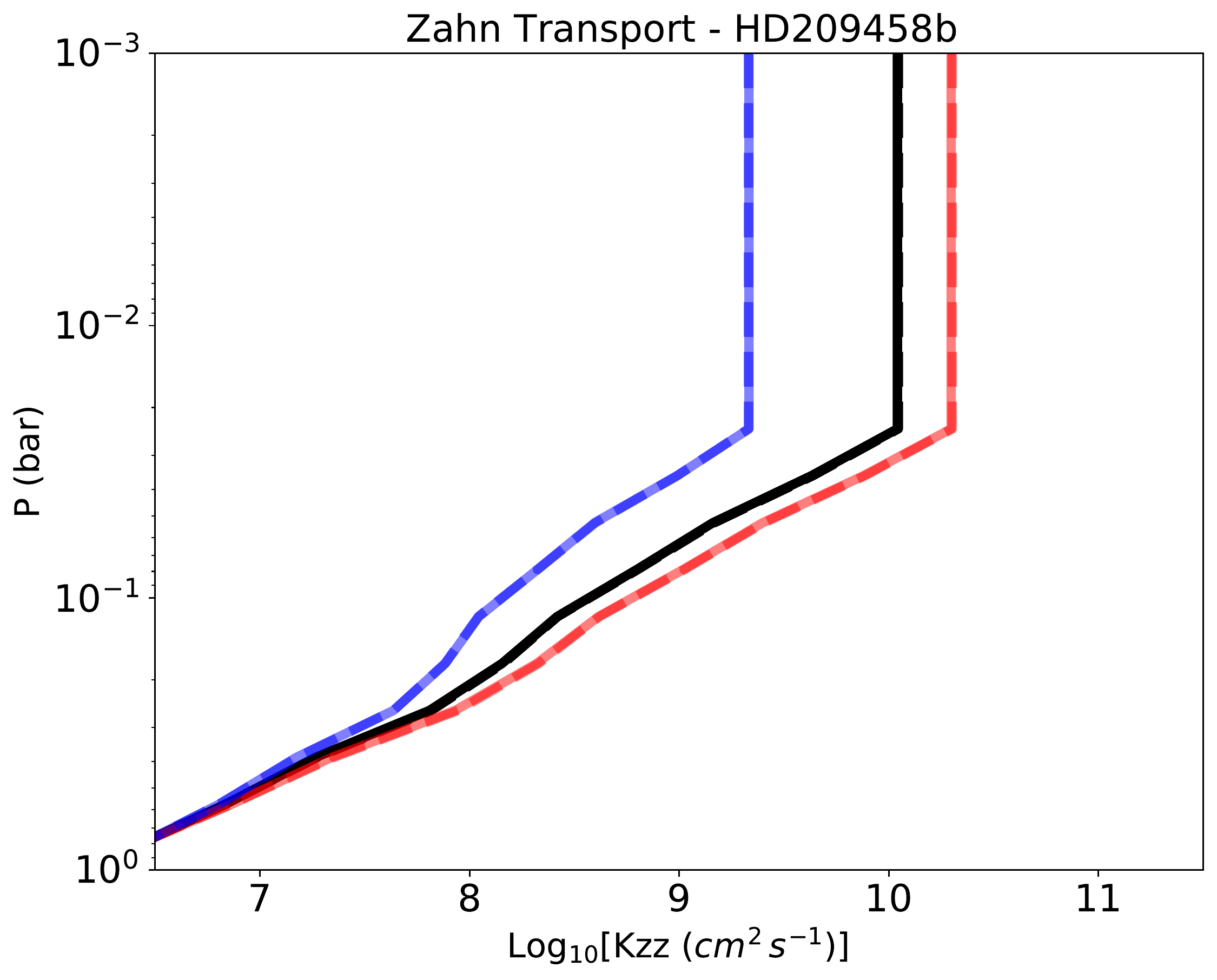}
	\includegraphics[width=\columnwidth]{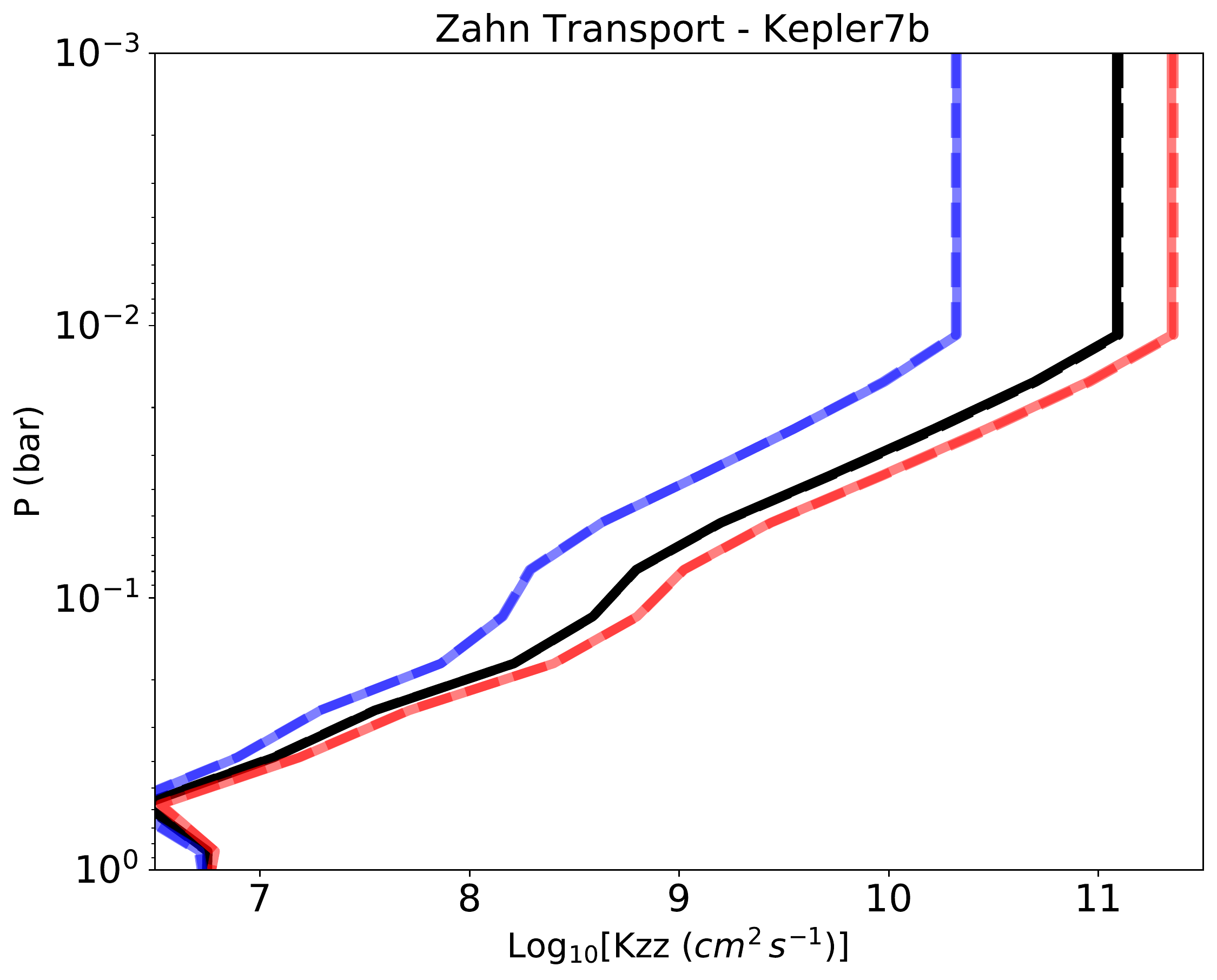}
    \caption{Same as Figure C1,  for the Zahn transport formulation. The differences between iterated and non-iterated profiles is barely visible in this case,  with a weaker vertical transport. }
    \label{fig:Kzz_Zahn_iterated}
\end{figure}

Figures~\ref{fig:Kzz_Max_iterated} and~\ref{fig:Kzz_Zahn_iterated} illustrate how diagnostic Kzz profiles obtained from atmospheric flows with or without feedback are very similar, for all three planets of interest.  This suggests that there is be little to negligible turbulent feedback on the mean atmospheric flow, as discussed in \S3.3 

\label{sec:feedback}

\section{Semi-Transparent Turbulence Properties}

We have argued in \S3.2.2 that the energy injection scale in an atmosphere subject to semi-transparent secular shear instabilities is the scale $\Lambda$ at which the medium transitions from being transparent to being opaque. Since this $\Lambda$ scale is smaller than the pressure-scale height below photospheric levels,  it is of interest to evaluate the dynamic range of the inertial turbulent cascade that will be permitted under such circumstances.

Borrowing some of the general scalings from \cite{2010ApJ...725.1146L}, we obtain an estimate of the  gas mean free path, scaled to $H_p$:
\begin{equation}
\frac{\lambda_{\rm mfp}}{H_p} \sim \frac{1}{5 \times 10^8} \left(  \frac{P}{{\rm mbar}}  \right)^{-1},
\end{equation}
which evaluates to $\sim 10^{-9}$ at photospheric levels.

From Kolmogorov turbulence scalings \citep{1987flme.book.....L}, the turbulent energy injection rate can be evaluated as
\begin{equation}
\epsilon = \frac{U^3}{L} = S^3 L^2 = S^3 \Lambda^2,
\end{equation}
where $U$ is a velocity scale,  $L$ is a lengthscale, $S$ is the shear rate and $\Lambda$ is the injection scale. Adopting a shear rate representative of hot Jupiter atmospheric flows,  $S \sim C_s / 2 H_p$ which evaluates to $\sim 3 \times 10^{-4}$s$^{-1}$, we obtain 
 \begin{equation}
\epsilon \simeq 10^{-2}  \frac{C_s^3}{H_p} \left(  \frac{\Lambda}{\Lambda _{\rm max}} \right)^2,
\end{equation}
where $C_s$ is the sound speed and $\Lambda _{\rm max} = 0.3 H_p$ is our adopted maximum integral scale. The turbulence dissipation scale is 
\begin{equation}
\eta_K = \left( \frac{\nu^3}{\epsilon} \right)^{1/4}   \simeq    10^{-7}  H_p \left( \frac{ \Lambda } {\Lambda_{\rm max}} \right)^{-1/2},
\end{equation}
where we have evaluated the kinematic viscosity $\nu$ as $ \sim C_s \lambda_{\rm mfp}$ and used our previous expression for $\lambda_{\rm mfp}$.

Inspecting Figure~\ref{fig:length}, we infer approximate inertial ranges for the turbulence at two pressure levels:
\begin{eqnarray}
 \sim \left[ 10^{-1},  10^{-7}\right] H_p& {\rm at }& 10^{-2}\,{\rm bar \, (photosphere)}\\
 \sim \left[ 10^{-3},  10^{-6}\right] H_p & {\rm at }& 1\,{\rm bar},
\end{eqnarray}
which shows that the inertial range quickly shrinks below photospheric levels.

We can also establish that rotation does not strongly affect the injection scale modes by considering their Rossby number, which is $Ro \sim S/\Omega \sim 10$ for a representative hot Jupiter rotation rate.

%%%%%%%%%%%%%%%%%%%%%%%%%%%%%%%%%%%%%%%%%%%%%%%%%%

% Don't change these lines
\bsp	% typesetting comment
\label{lastpage}
\end{document}